\documentclass[epj,final,numComoabook]{svjour} 
\usepackage{graphics}
\usepackage{multirow} 
\usepackage{amsmath} 
\usepackage{amssymb}
\usepackage{epsfig} 
\usepackage{graphicx} 
\usepackage{color}
\usepackage{xspace} 
\usepackage[colorlinks=true,
  pdfstartview=FitV,linkcolor=blue,citecolor=blue,urlcolor=blue]{hyperref}
\newcommand{\dAu}{\mbox{${\rm {\it d}+Au}$}\xspace}
\newcommand{\pAl}{\mbox{$p+{\rm Al}$}\xspace}
\newcommand{\pAu}{\mbox{$p+{\rm Au}$}\xspace}

\newcommand{\HeAu}{\mbox{${\rm ^{3}He+Au}$}\xspace}
\newcommand{\AuAu}{\mbox{${\rm Au+Au}$}\xspace}
\newcommand{\UU}{\mbox{${\rm U+U}$}\xspace}
\newcommand{\PbPb}{\mbox{${\rm Pb+Pb}$}\xspace}
\newcommand{\pPb}{\mbox{${\rm {\it p}+Pb}$}\xspace}
\newcommand{\CuCu}{\mbox{${\rm Cu+Cu}$}\xspace}
\newcommand{\CuAu}{\mbox{${\rm Cu+Au}$}\xspace}
\newcommand{\pp}{\mbox{${p+p}$}\xspace}
\newcommand{\AAA}{\mbox{${\rm A+A}$}\xspace}

\newcommand{\ee}{\mbox{${e^{+}e^{-}}$}\xspace}
\newcommand{\pA}{\mbox{$p+{\rm A}$}\xspace}

\newcommand{\pT}{\mbox{${p_{\rm{_{T}}}}$}\xspace}
\newcommand{\vell}{\mbox{${v_{2}}$}\xspace}
\newcommand{\Nch}{\mbox{${\rm N_{ch}}$}\xspace}
\newcommand{\nvq}{\mbox{${\rm n_{vq}}$}\xspace}
\newcommand{\ET}{\mbox{${\rm E_{T}}$}\xspace}

\newcommand{\Jpsi}{\mbox{${J/\psi}$}\xspace}

\newcommand{\Ncoll}{\mbox{$N_{\rm coll}$}\xspace}
\newcommand{\dNch}{\mbox{${\rm dN_{ch}/d\eta}$}\xspace}

\newcommand{\snn}{\mbox{$\sqrt{s_{_{NN}}}$}\xspace}
\newcommand{\s}{\mbox{$\sqrt{s}$}\xspace}

\newcommand{\dnch}{\mbox{${\rm dN_{ch}/d\eta}$}\xspace}
\newcommand{\Tch}{\mbox{${\rm T_{ch}}$}\xspace}
\newcommand{\mb}{\mbox{${\mu_{\rm B}}$}\xspace}
\newcommand{\ms}{\mbox{${\mu_{\rm S}}$}\xspace}
\newcommand{\mq}{\mbox{${\mu_{\rm Q}}$}\xspace}

\newcommand{\avgNp}{\mbox{${\rm \langle N_{part} \rangle}$}\xspace}
\newcommand{\Np}{\mbox{${\rm N_{part}}$}\xspace}
\newcommand{\Nc}{\mbox{${\rm N_{coll}}$}\xspace}
\newcommand{\bim}{\mbox{${b}$}\xspace}
\newcommand{\avgb}{\mbox{$\langle b \rangle$}\xspace}

\newcommand{\pt}{\mbox{$p_{\rm{T}}$}\xspace}

\newcommand{\KET}{\mbox{${\rm KE_{\rm T}}$}\xspace}
\newcommand{\RAA}{\mbox{${R_{AA}(\pT)}$}\xspace}
\newcommand{\pta}{\mbox{${\rm p^{a}_{_{T}}}$}\xspace}
\newcommand{\ptt}{\mbox{${\rm p^{t}_{_{T}}}$}\xspace}
\newcommand{\lnsnn}{\mbox{${\rm ln(\sqrt{s_{_{NN}}})}$}\xspace}
\newcommand{\sn}{\mbox{${\rm \sqrt{s_{_{NN}}}}$}\xspace}
\newcommand{\Nsp}{\mbox{${\rm  N_{spec} }$}\xspace}
\newcommand{\dnchmid}{\mbox{${\rm dN_{ch}/d\eta|_{|\eta| <1}/\langle N_{part}/2 \rangle}$}\xspace}
\newcommand{\dnchNp}{\mbox{${\rm dN_{ch}/d\eta/\langle N_{part}/2 \rangle}$}\xspace}

\def\FigurePhase{
\begin{figure}
\begin{center}
\resizebox{0.5\textwidth}{!}{\includegraphics{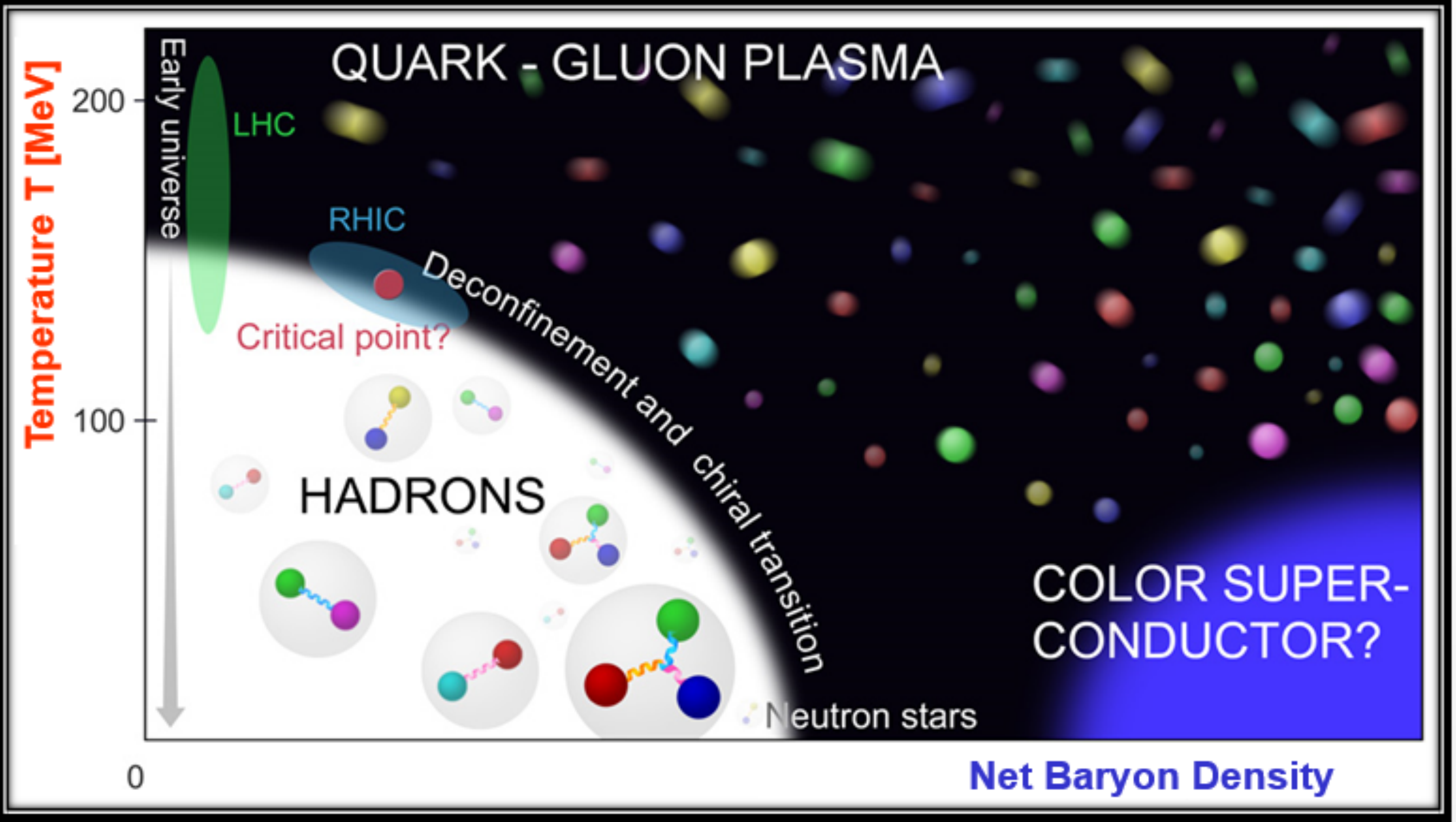}}
\caption{The phase diagram of QCD matter showing the possible states
  of quarks. Ordinary matter lies at the bottom left corner, where the
  temperature and the density of matter both are relatively low.
  Researchers at the RHIC and LHC explore much higher energies;
  currently RHIC's researchers are hunting for the possible existence
  of a critical point at which stable hadrons may mingle with the
  quark-gluon plasma. This figure is adapted from Ref.~\cite{phase}.
\label{fig:Phase}}
\end{center}
\end{figure}
}

\def\FigureQCD{
\begin{figure}
\begin{center}
\resizebox{0.5\textwidth}{!}
{\includegraphics{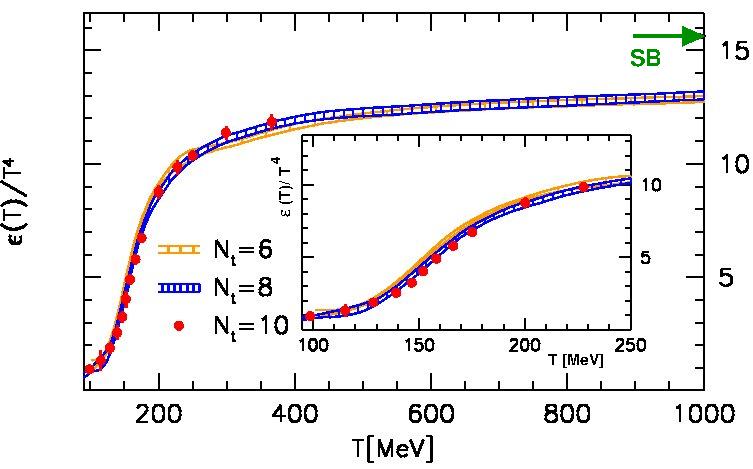}}
\caption{The energy density normalized by \mbox{${\rm T^{4}}$} as a
  function of the temperature on \mbox{$N_{t} = $ 6, 8 and 10}
  lattices. An arrow indicates the Stefan-Boltzmann limit
  $\varepsilon_{_{SB}}$ = 3 $p_{_{SB}}$ (see text for details). This figure is adapted from
  Ref.~\cite{Sza2010}.}
\label{fig:QCD} 
\end{center}
\end{figure}
}

\def\FigureSoftpeak{
\begin{figure}
\begin{center}
\resizebox{0.5\textwidth}{!}{\includegraphics{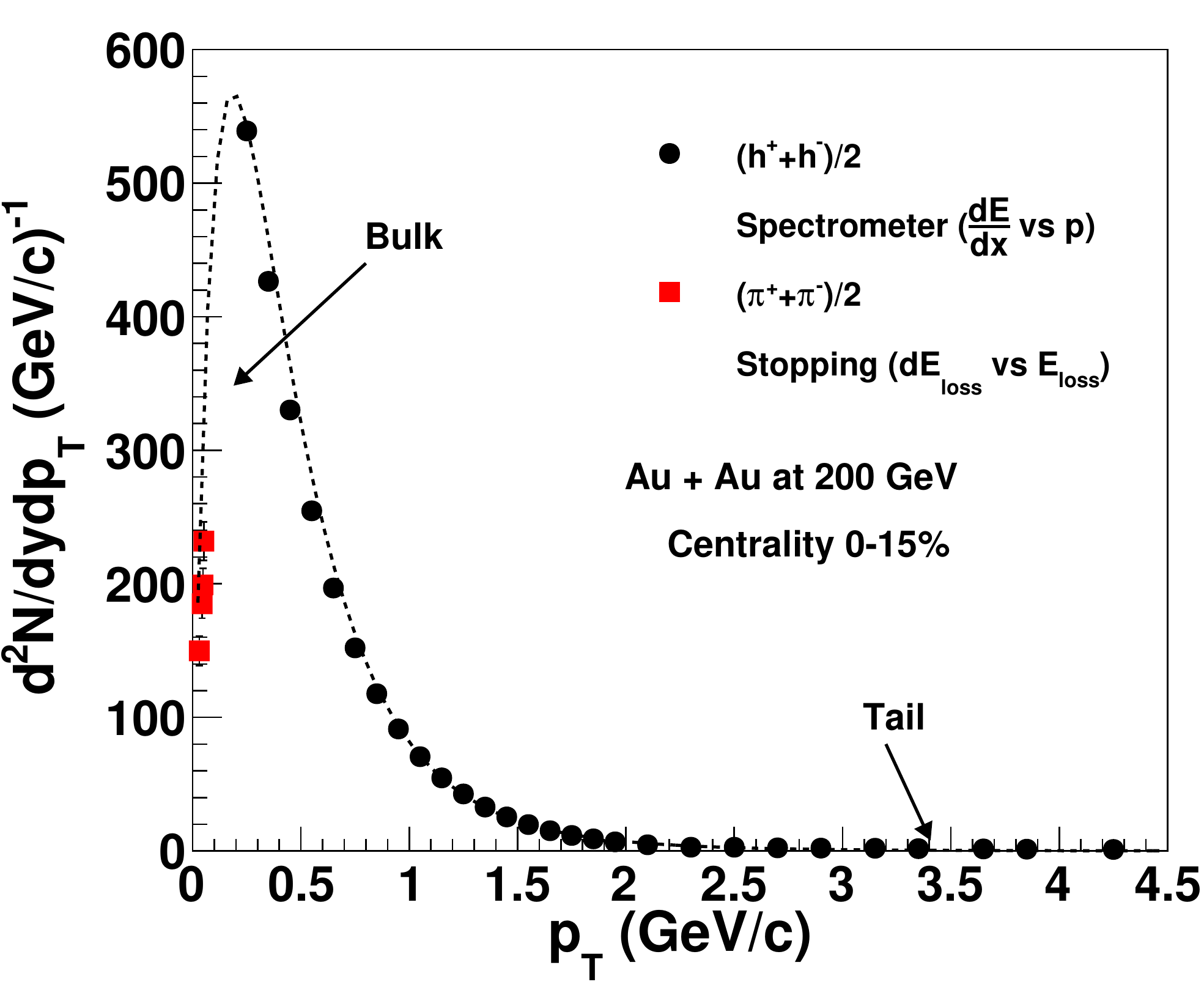}}
\caption{The measured distribution of charged hadrons as a function of
  transverse momentum, \pT, for the 0-15$\%$ most central \AuAu
  collisions at \snn = 200 GeV \cite{Rachid2003}.}
\label{fig:Softpeak} 
\end{center}
\end{figure}
}

\def\FigureGlauber{
\begin{figure}
\begin{center}
\resizebox{0.47\textwidth}{!}{\includegraphics{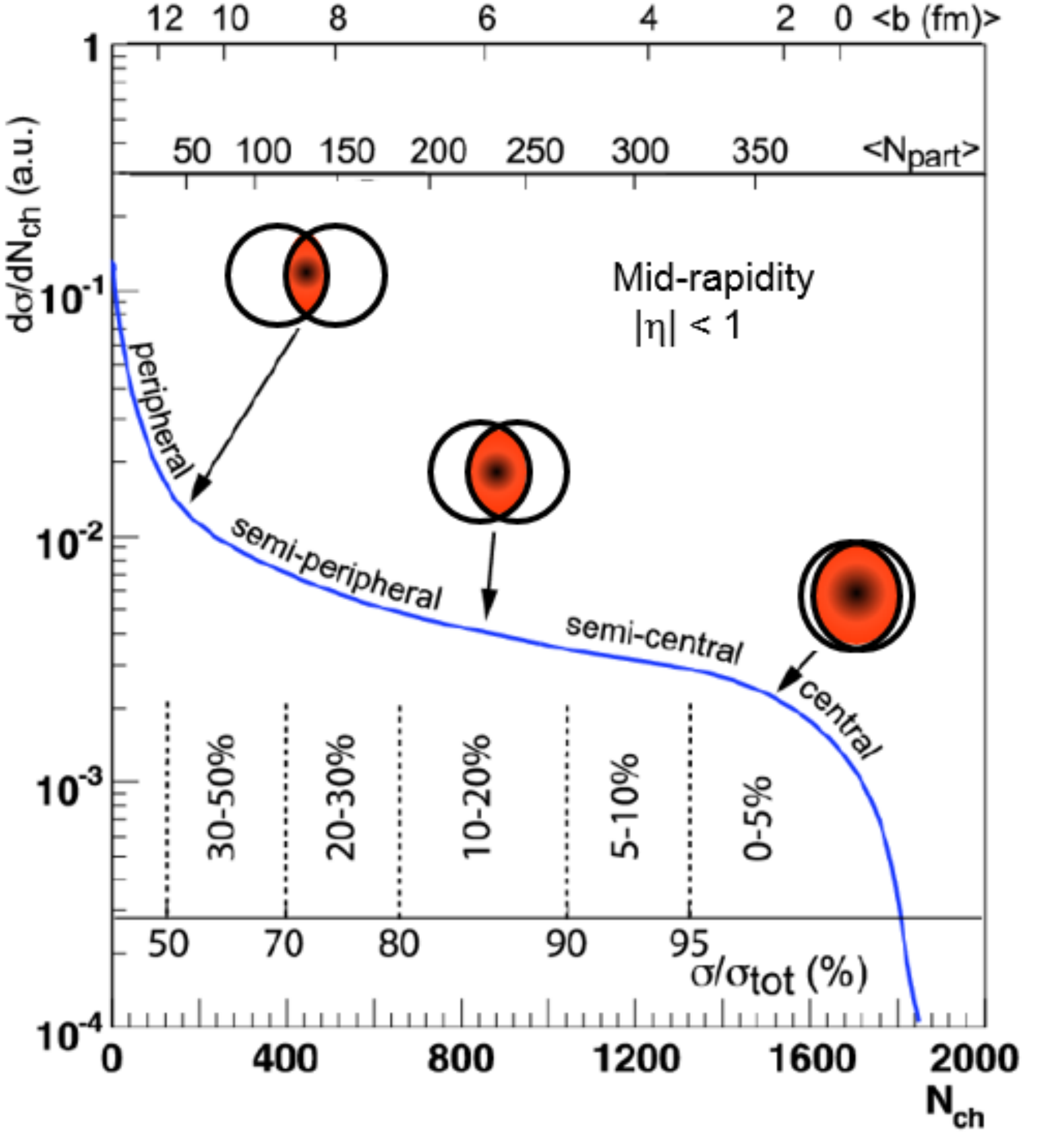}}
\caption{Schematic diagram of a distribution of a centrality
  variable, i.e. the number of charged particles in the interval
  $|\eta| <$ 1. The events contributing to the upper 5\% of the
  integral of the distribution are the 0-5\% centrality bin, with a
  near complete nuclear overlap. The parameters from a Glauber
  simulation, \avgb, \avgNp, are shown as different horizontal
  scales. This figure is adapted from Ref. \cite{Mich2007}.}
\label{fig:Glauber} 
\end{center}
\end{figure}
}

\def\FigureMultMid{
\begin{figure}
\begin{center}
\resizebox{0.465\textwidth}{!}{\includegraphics{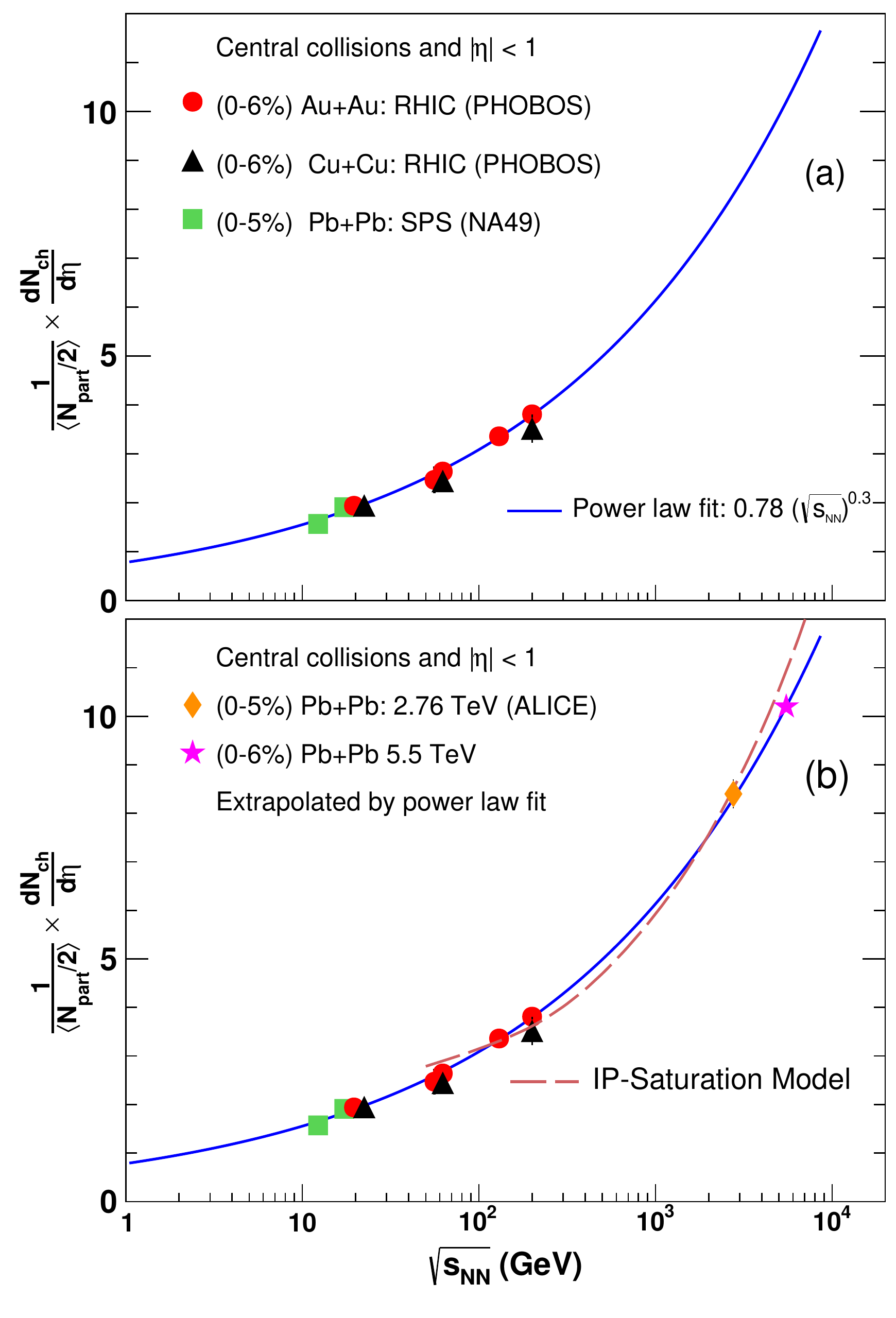}}
\caption{The measured scaled pseudorapidity density ${\rm
    dN_{ch}/d\eta /\langle {1\over2}N_{part}\rangle}$ for ${\rm \mid
    \eta \mid < 1}$ in central \AuAu, \CuCu and \PbPb collisions at the
  SPS, RHIC and LHC energies
  \cite{RachidMoriond2002,RachidHDR,AuAufrag,Panic2006,CuCu2008,SPSdNdEta1,SPSdNdEta2,ALICEPb276}. The
  star symbols denote the extrapolation in central \PbPb collisions at
  5.5 TeV (LHC). The error bars show the systematic errors.  The solid
  curve corresponds to the power law fit of the RHIC data. The Impact
  Parameter dipole saturation model (IP-saturation model) calculation
  from RHIC to LHC is illustrated by a dashed curve \cite{IPCGC2012}. See text for details}
\label{fig:MultMid}
\end{center}
\end{figure}
}

\def\FigureMult4pi{
\begin{figure*}
\begin{center}
  \resizebox{\textwidth}{!}{\includegraphics{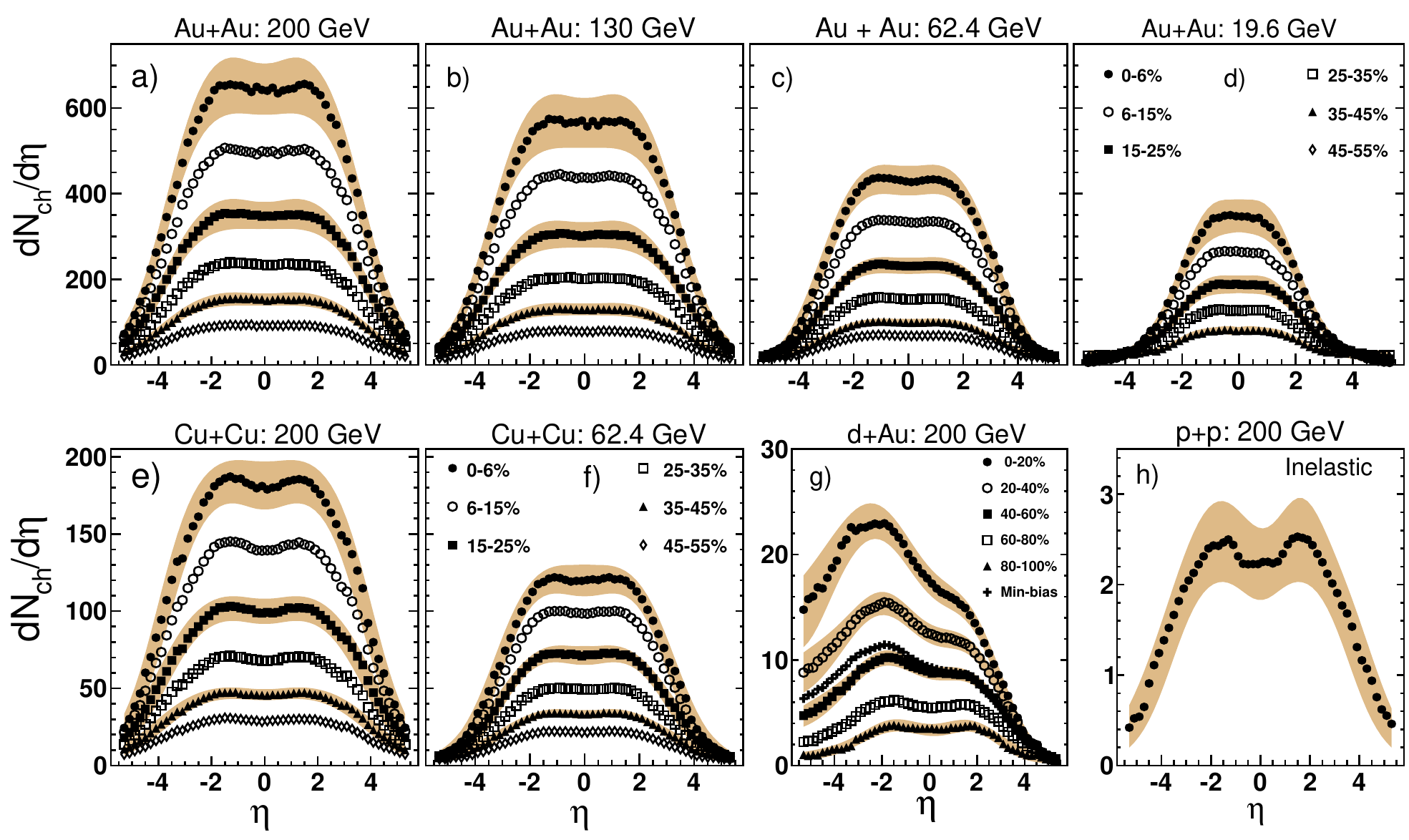}}
\caption{The measured pseudorapidity density distributions of charged
  particles produced in \AuAu, \CuCu, \dAu, and \pp collisions at RHIC
  energies. The \dNch\ distributions for \AuAu, \CuCu and \dAu
  collisions are plotted as a function of the collision's
  centrality. Typical systematic errors are represented as bands for
  selected centrality bins. The statistical errors are negligible
  \cite{RachidHDR,RachidMoriond2002,AuAufrag,Panic2006,CuCu2008}.}
\label{fig:Mult4pi} 
\end{center}
\end{figure*}
}

\def\FiguredNdEtaCuCuAuAu{
\begin{figure*}
\begin{center}
\resizebox{0.72\textwidth}{!}
{\includegraphics{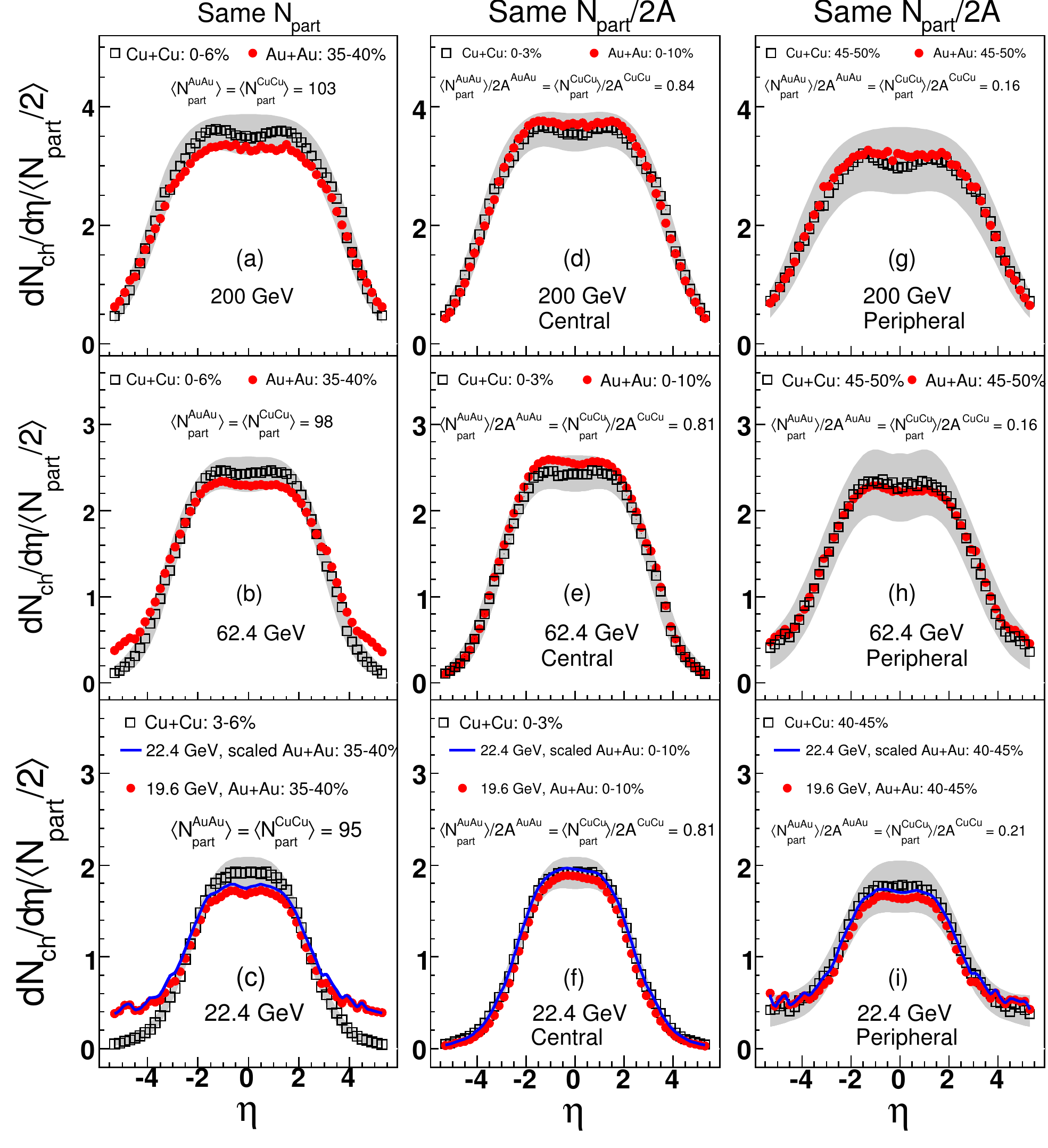}}
\caption{The measured \dnchNp\ distributions in \CuCu and \AuAu
  collisions at 22.4 (19.6), 62.4 and 200 GeV selected to yield the
  following: Panels (a)-(c) similar \avgNp, panels (d)-(f) central
  collisions with similar volume of overlap region, \avgNp/2A and
  panels (g)-(i) peripheral collisions with similar volume of overlap
  region (i.e., the fraction of the total nuclear volume that
  interacts), \avgNp/2A . The band indicates the systematic
  uncertainty for \CuCu collisions. For the sake of clarity, the
  errors for \AuAu are not shown
  \cite{RachidHDR,RachidMoriond2002,AuAufrag,Panic2006,CuCu2008}.}
\label{fig:dNdEtaCuCuAuAu} 
\end{center}
\end{figure*}
}

\def\FigureRatioBrahms{
\begin{figure}
\begin{center}
\resizebox{0.47\textwidth}{!}{\includegraphics{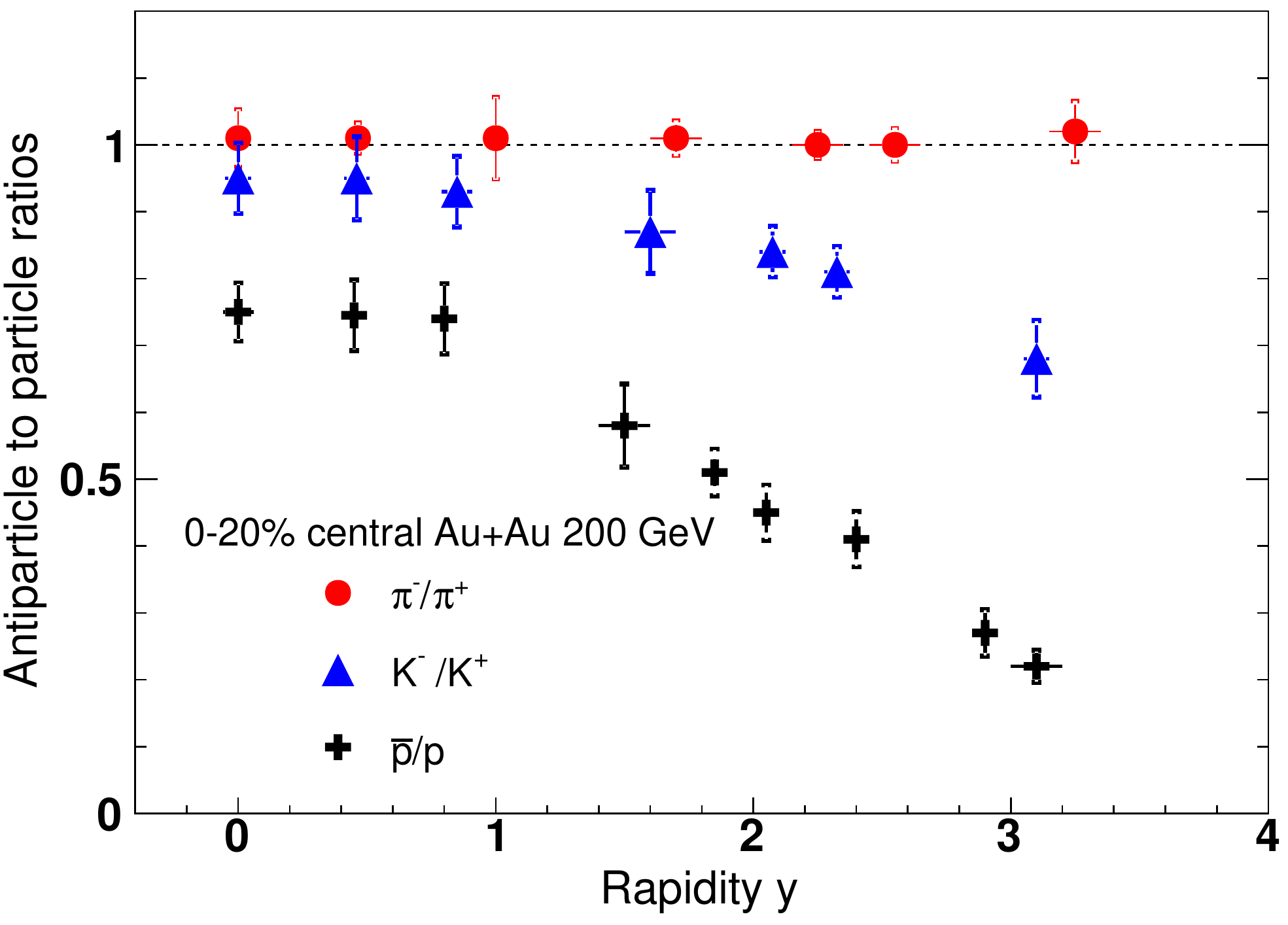}}
\caption{The measured antiparticle to particle ratios as a function of
  rapidity in 0-20\% central \AuAu collisions at \snn~=~200~GeV. Error
  bars show the statistical errors while the caps indicate the
  combined statistical and systematic errors~\cite{BRAHMSRatio}.}
\label{fig:RatioBrahms} 
\end{center}
\end{figure}
}

\def\FigureRatiotherm{
\begin{figure}
\begin{center}
\resizebox{0.47\textwidth}{!}{\includegraphics{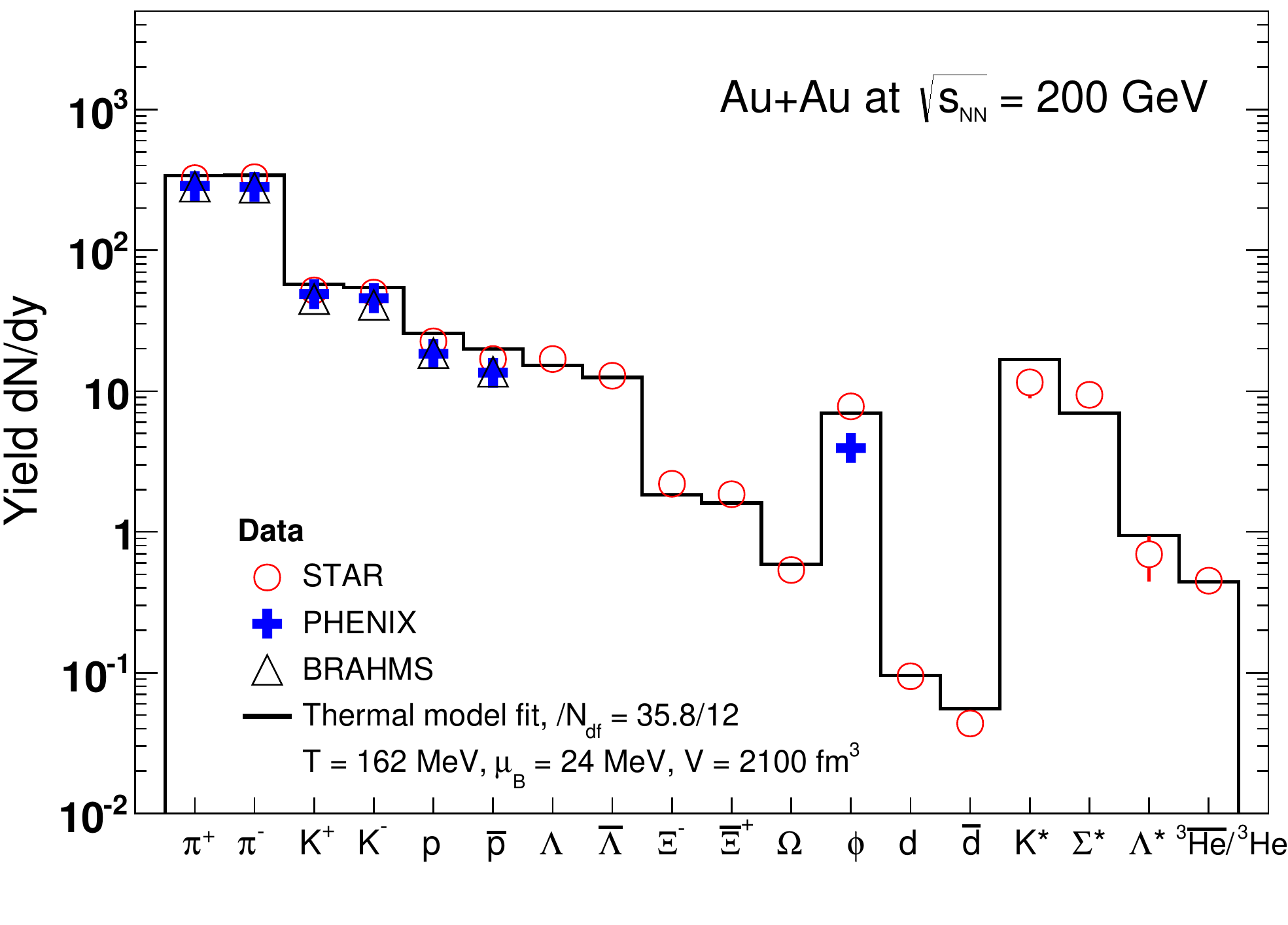}}
\caption{The experimental hadron yields and statistical thermal model
  comparison for different particle species produced in central \AuAu
  at \snn = 200 GeV~\cite{Anton2013}.}
\label{fig:Ratiotherm} 
\end{center}
\end{figure}
}

\def\FigureRatioStar{
\begin{figure}
\begin{center}
\resizebox{0.45\textwidth}{!}{\includegraphics{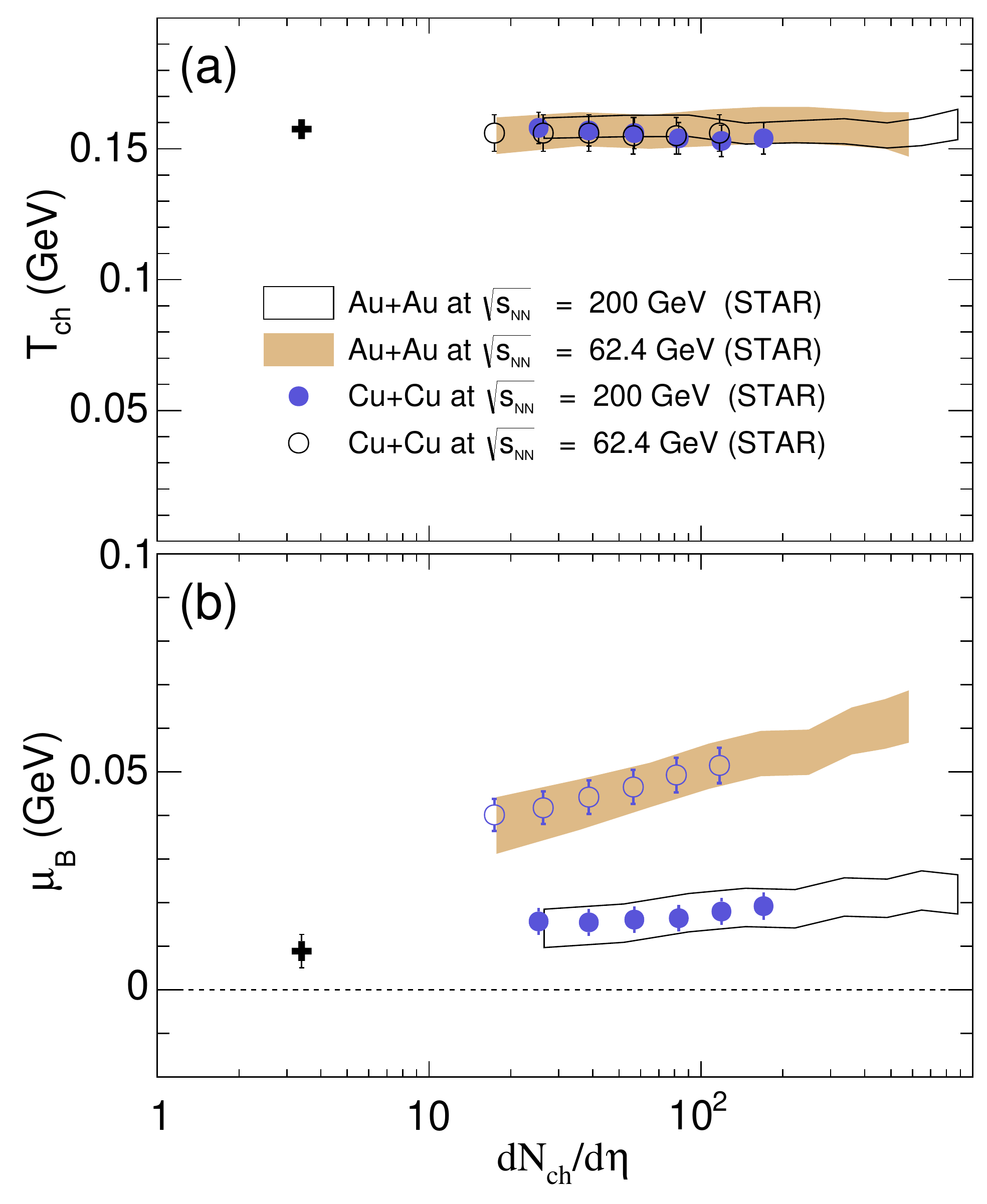}}
\caption{The chemical freeze-out temperature, \Tch, and baryon
  chemical potential, \mb, as a function of centrality \dnch shown
  respectively in panels (a) and (b). The data are for \snn = 200 and
  62.4 GeV in \AuAu (bands) and \CuCu (symbols). For comparison, the
  results for minimum-bias \pp collisions at \snn = 200 GeV are also
  shown using data from Ref. \cite{StarPRC2011}.}
\label{fig:RatioStar} 
\end{center}
\end{figure}
}

\def\FigureRatioLEP{
\begin{figure}
\begin{center}
\resizebox{0.5\textwidth}{!}{\includegraphics{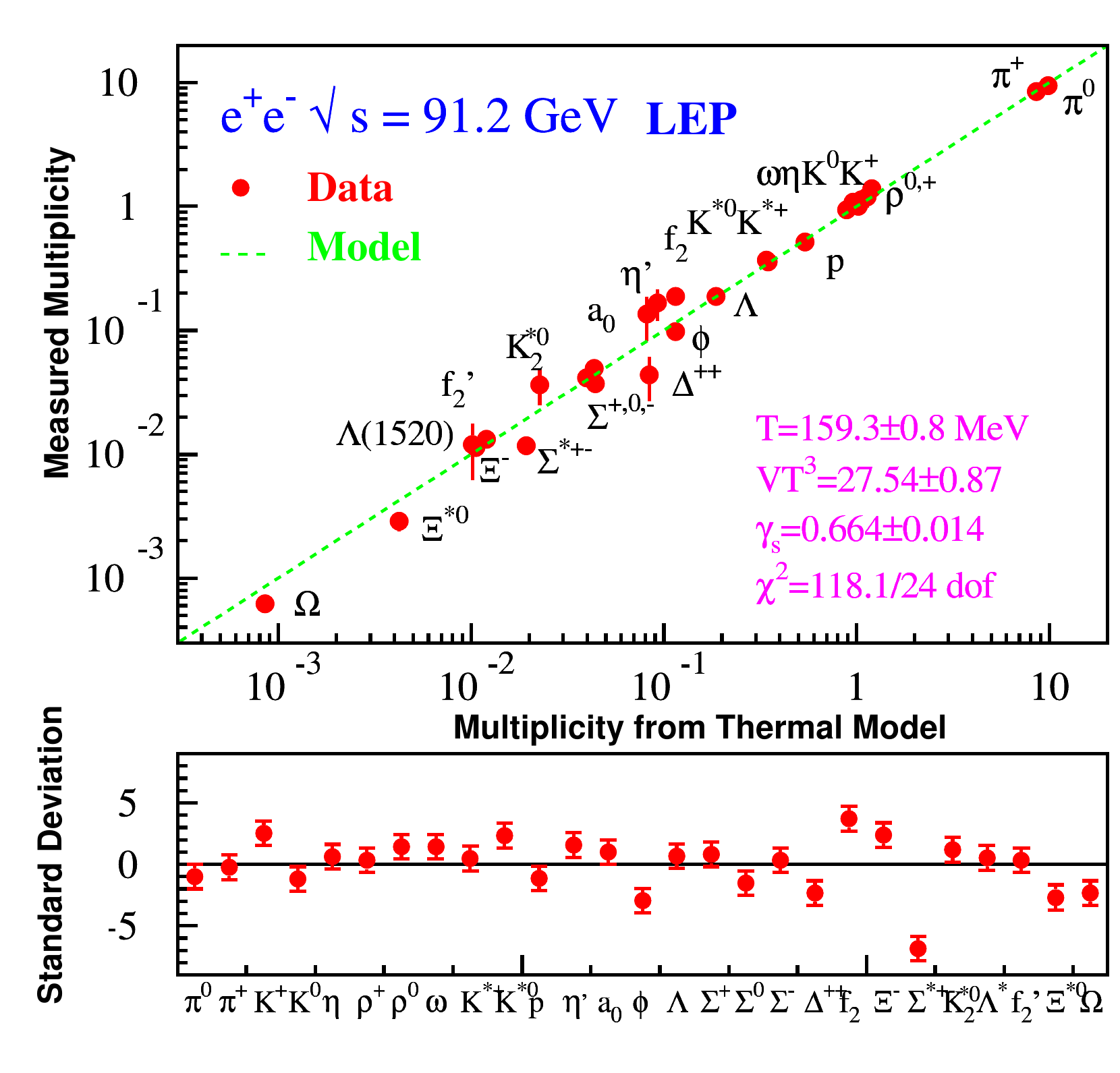}}
\caption{Experimental hadron yields and statistical thermal
  model comparison for different particle species produced in LEP \ee
  annihilation at \s = 91.2 GeV. This figure is adapted from
  Ref. \cite{Klim2010}.}
\label{fig:RatioLEP}
\end{center}
\end{figure}
}

\def\FigureVSch{
\begin{figure*}
\begin{center}
\resizebox{0.8\textwidth}{!}{\includegraphics{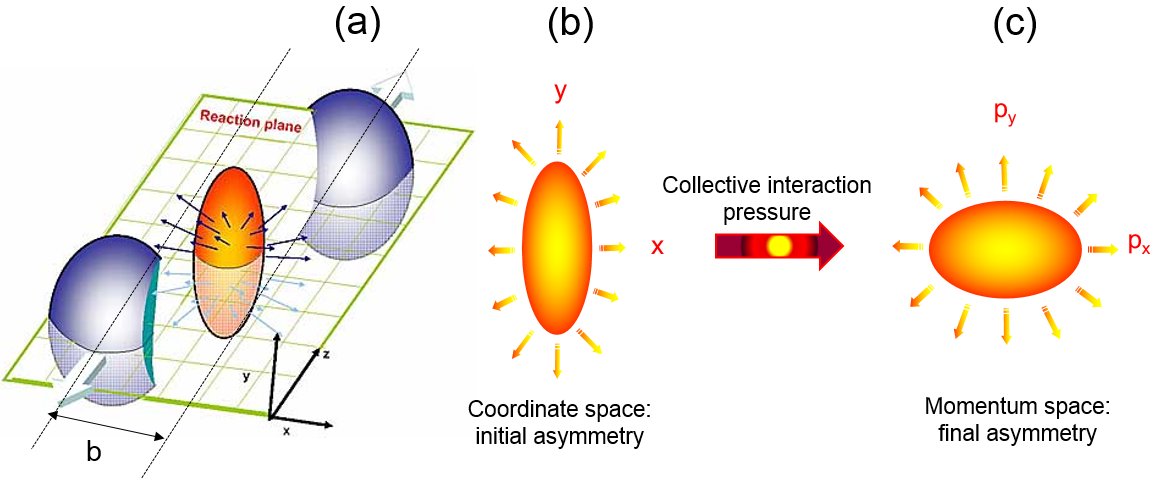}}
\caption{Panel (a) presents schematically the collision zone between
  two incoming nuclei. Panel (b) illustrates initial-state anisotropy
  in the collision zone converting into final-state elliptic flow, and
  panel (c) measured as anisotropy in particle momentum.}
\label{fig:VSch} 
\end{center}
\end{figure*}
}

\def\FigureVEta{
\begin{figure*}
\begin{center}
\resizebox{1.\textwidth}{!}{\includegraphics{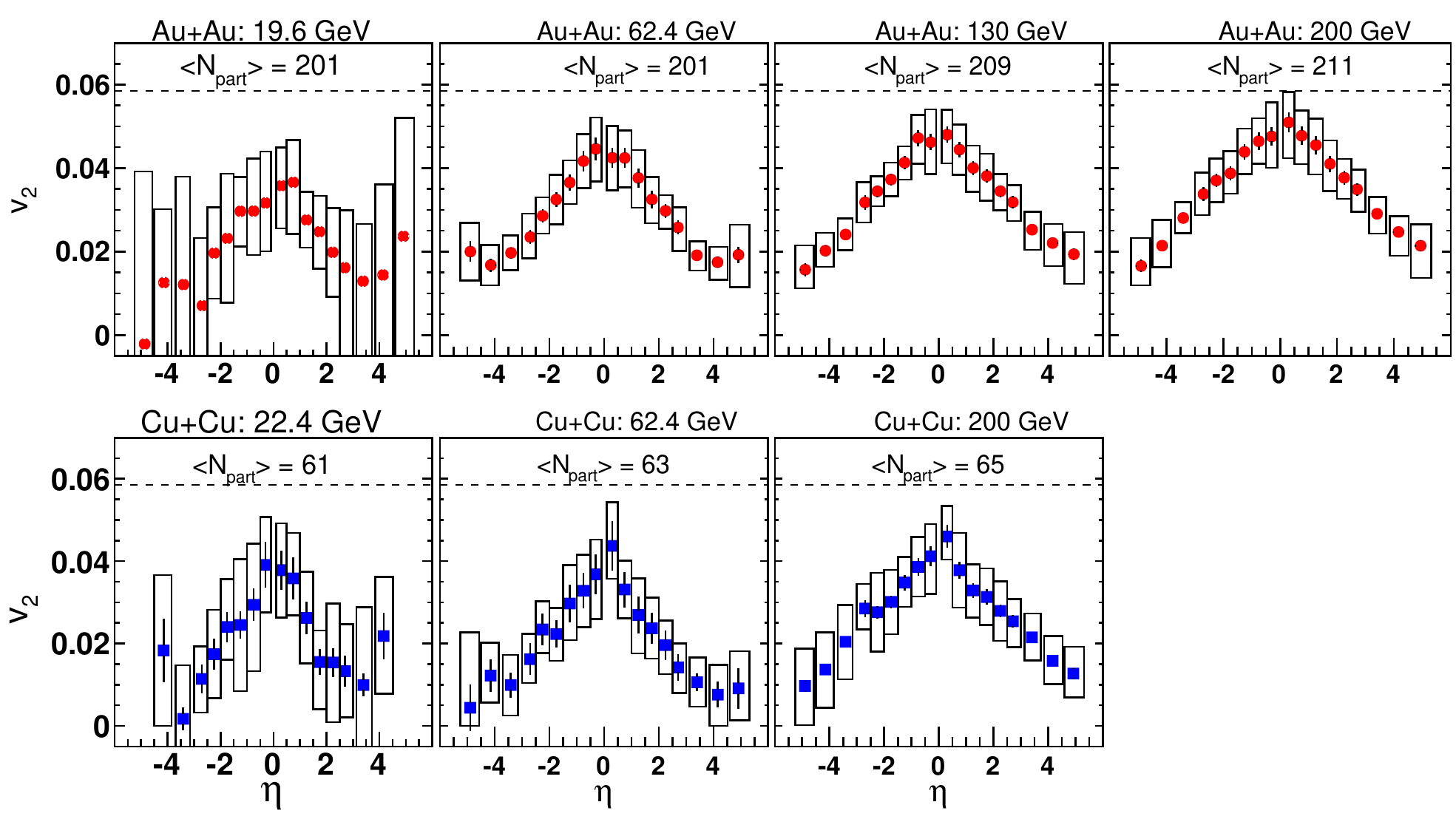}}
\caption{The measured elliptic flow, \vell, vs $\eta$ for \AuAu and
  \CuCu collisions at several RHIC energies.The boxes show the
  systematic errors and the bars represent the statistical errors
  \cite{RachidHDR,RachidQM2006,PHOflow1,PHOflow2}.}
\label{fig:VEta} 
\end{center}
\end{figure*}
}

\def\FigureVEngy{
\begin{figure}
\begin{center}
\resizebox{0.455\textwidth}{!}{\includegraphics{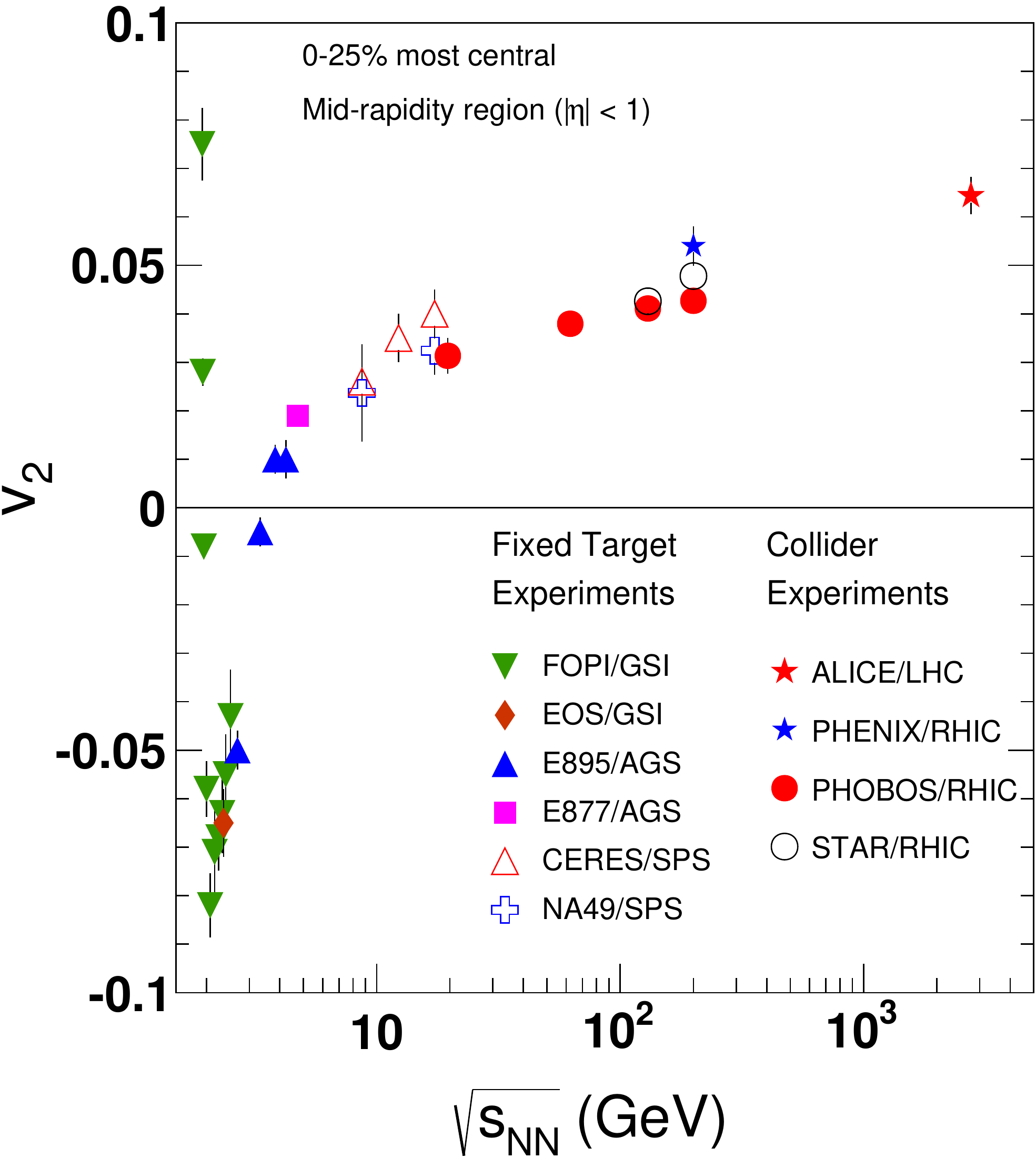}}
\caption{Compilation of data using
  Refs.~\cite{FOPI2005,RachidLK2007,Urs2008,Raimand2011_1,Raimand2011_2} of the dependence of
  integrated elliptic flow,\vell, on beam energy at GSI (\AuAu), AGS
  (\AuAu), SPS (\PbPb) and LHC (\PbPb).}
\label{fig:VEngy} 
\end{center}
\end{figure}
}

\def\FigureVpt{
\begin{figure}
\begin{center}
\resizebox{0.47\textwidth}{!}{\includegraphics{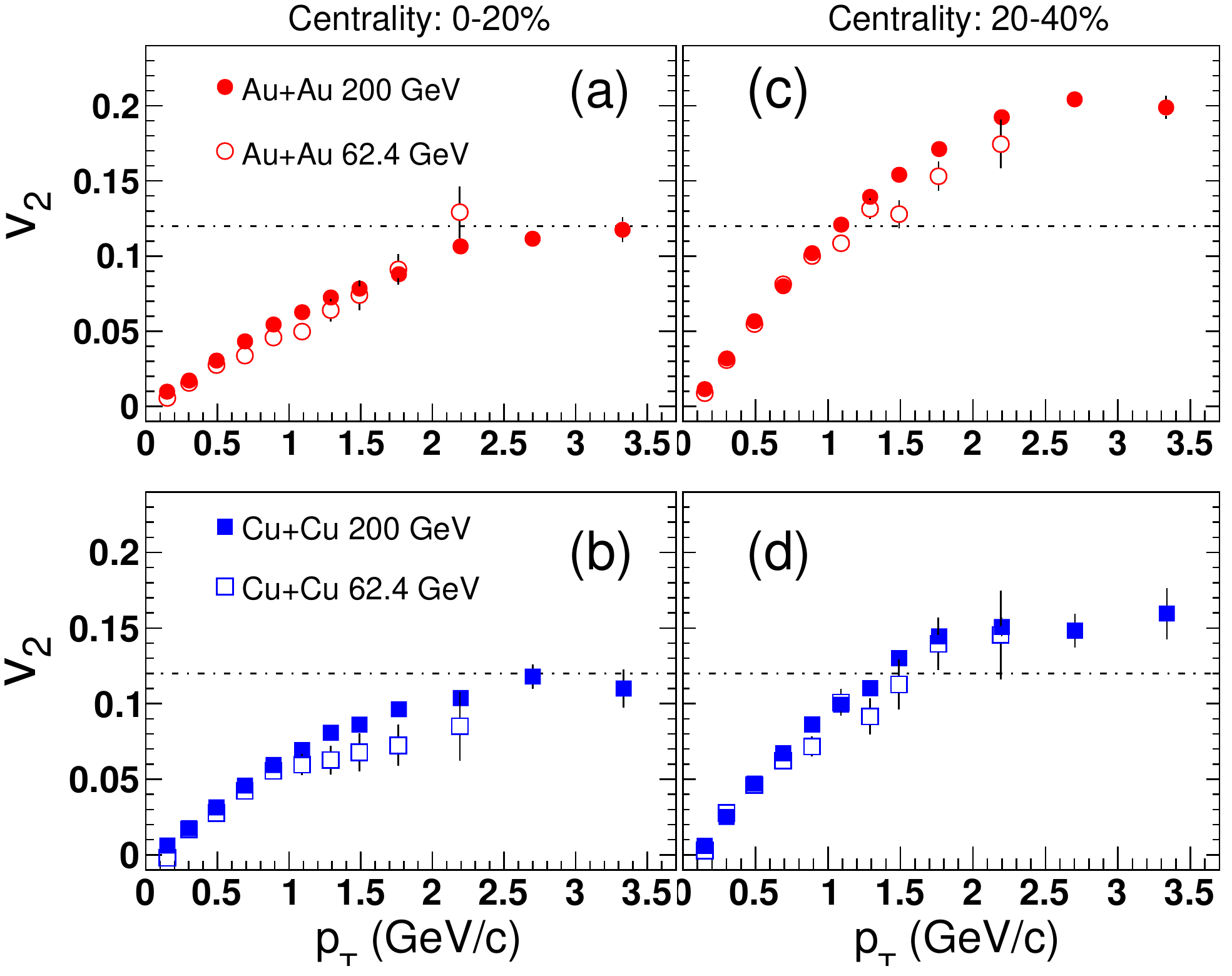}}
\caption{The measured elliptic flow, \vell, vs \pT for \AuAu and \CuCu
  collisions at several RHIC energies for centrality bins 0-20\% in panels (a)-(b)
  and 20-40\% in panels (c)-(d). The boxes show the systematic
  errors and the bars represent the statistical errors
  \cite{RachidHDR,RachidQM2006,PHOflow1,PHOflow2}.}
\label{fig:Vpt}
\end{center}
\end{figure}
}
\def\FigureVnqStar{
\begin{figure*}
\begin{center}
\resizebox{0.9\textwidth}{!}
{\includegraphics{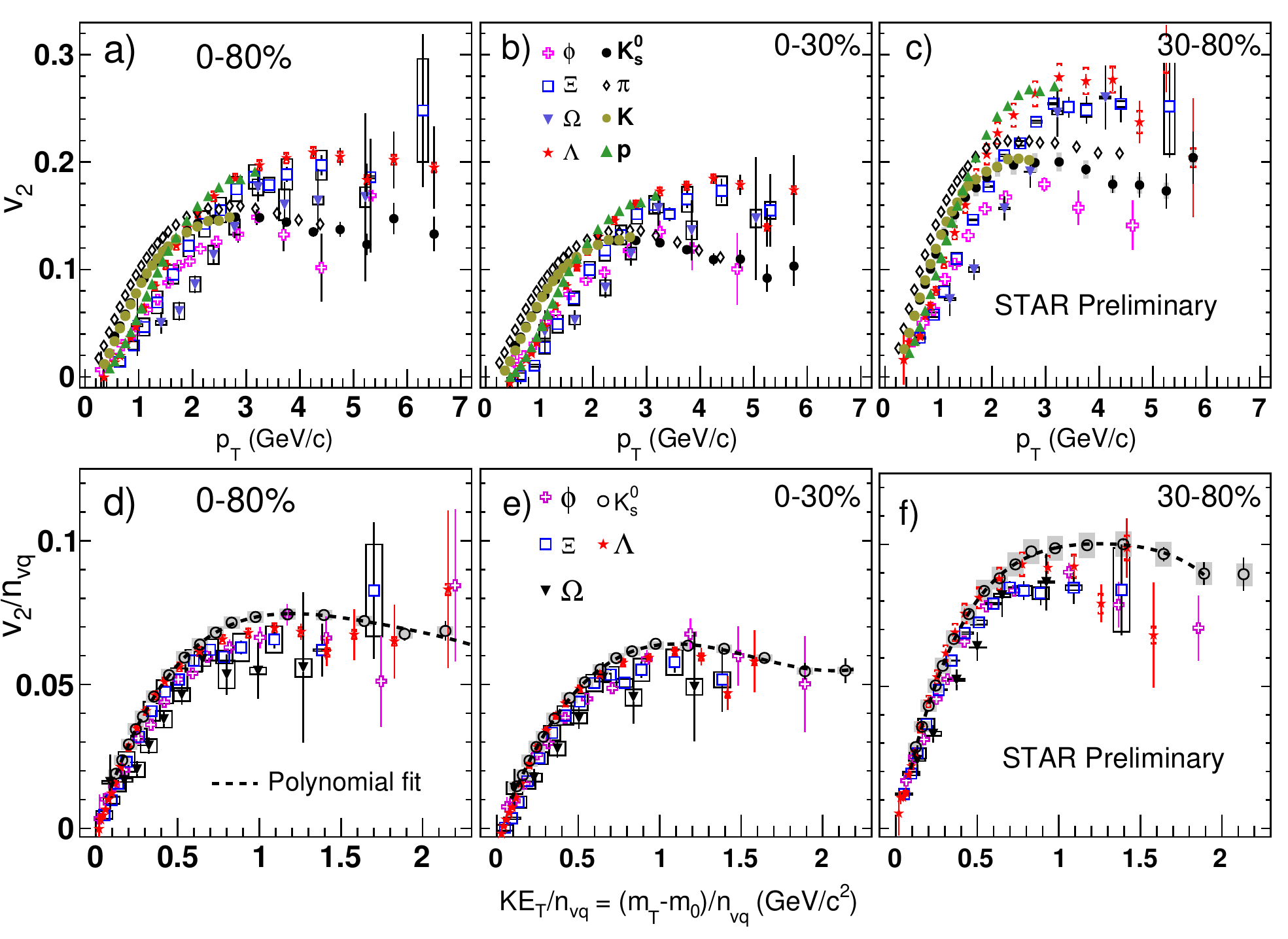}}
\caption{Panels (a), (b) and (c): the identified hadron anisotropy,
  \vell, as a function of \pT. Panels (d),
  (e) and (f): scaled identified hadron anisotropy, \vell/\nvq as a
  function of scaled transverse kinetic energy, \KET/\nvq.  \nvq is the number of valence quarks in a
  given hadron (for mesons, \nvq = 2; and, for baryons: 3).  All data
  are from \AuAu collisions at \snn = 200 GeV.  This figure is adapted
  from Ref.~\cite{Nasim2013}.}
\label{fig:VnqStar} 
\end{center}
\end{figure*}
}

\def\FigureVnqPhenix{
\begin{figure}
\begin{center}
\resizebox{0.45\textwidth}{!}
          {\includegraphics{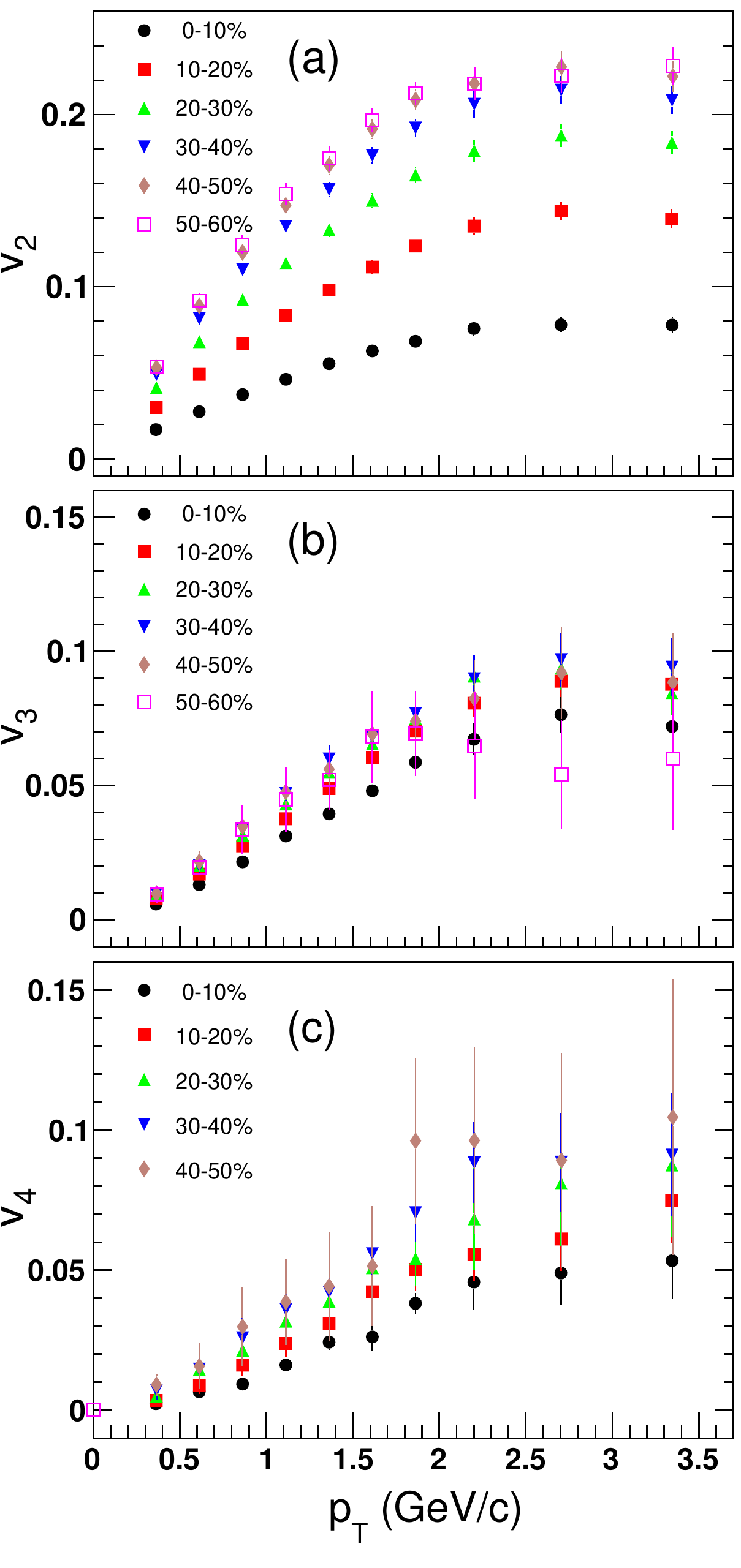}}
\caption{The measured harmonic flow $v_{n}$ ($n$ = 2, 3 and 4) versus
  \pt measured in various centralities in \AuAu at \mbox{\snn = 200
    GeV}. Error bars show combined statistical and systematic
  errors. Figure is made using data from Ref. \cite{ppg132}.}
\label{fig:VnqPhenix} 
\end{center}
\end{figure}
}

\def\FigureVnIP{
\begin{figure}
\begin{center}
\resizebox{0.5\textwidth}{!}
{\includegraphics{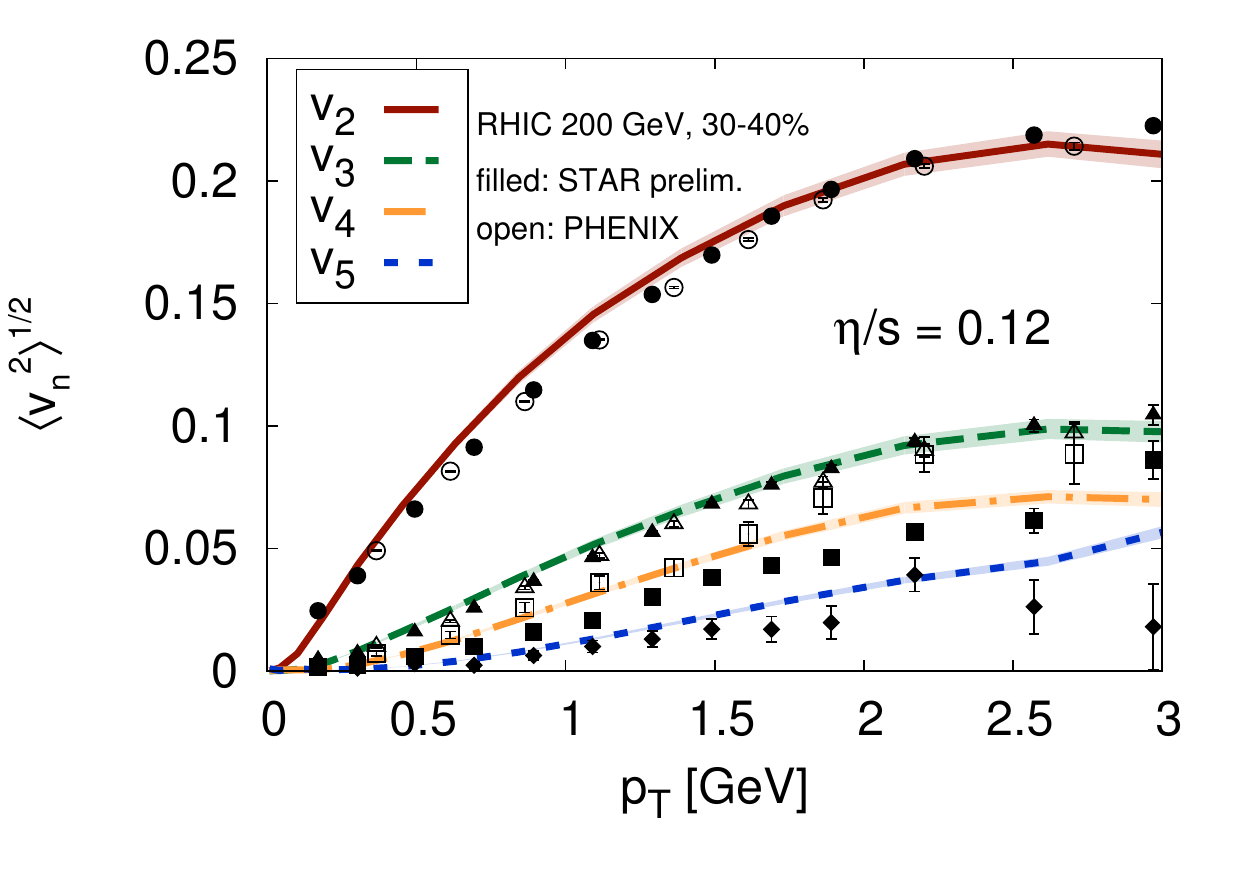}}
\caption{The comparison of the RMS anisotropic flow coefficient,
  $\langle v^{2}_{n}(\pT) \rangle^{2}$, as a function of \pT\ measured in \AuAu at 200 GeV at RHIC to IP-Glasma model using constant shear viscosity to entropy
  density ratio, $\eta/s$ $\approx$ 0.12. This figure is adapted from
  Ref. \cite{Char2013}.}
\label{fig:VnIP} 
\end{center}
\end{figure}
}

\def\FigureSpectra{
\begin{figure}
\begin{center}
\resizebox{0.45\textwidth}{!}
{\includegraphics{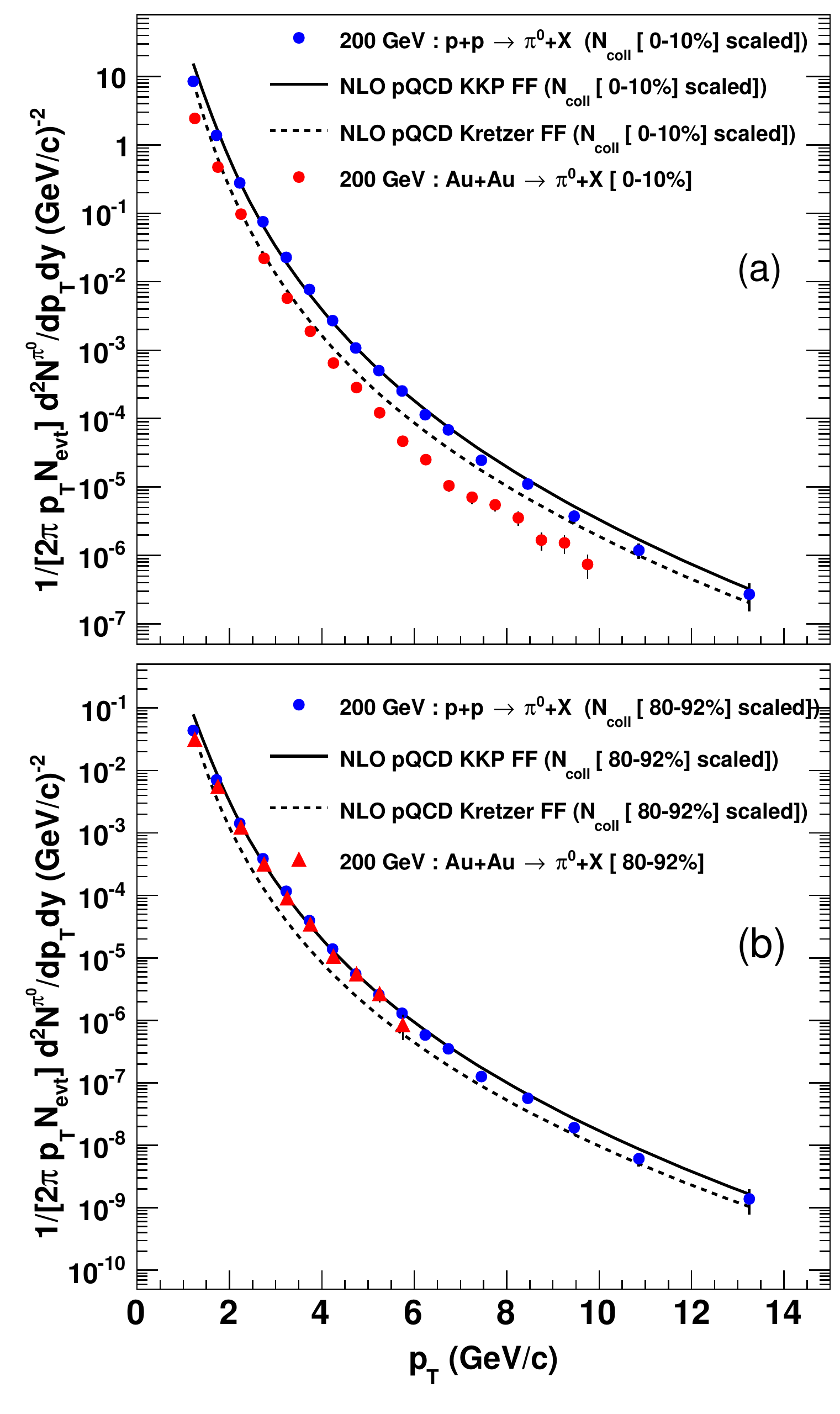}}
\caption{Invariant differential cross section for inclusive $\pi^{0}$
  production (points) and the results from NLO pQCD calculations with
  the same scaling using the ‘‘Kniehl-Kramer-Potter’’ (solid line) and
  ‘‘Kretzer’’ (dashed line) sets of fragmentation functions. The \pp
  data are scaled by the number of binary collisions, \Ncoll,
  corresponding to the centrality class in \AuAu collisions: panel (a) 0-10\% and panel (b) 80-92\%. Data used
  in the figure were taken from Refs. \cite{ppg014,ppg024}.}
\label{fig:Spectra} 
\end{center}
\end{figure}
} 

\def\FigureRaaSPSLHC{
\begin{figure}
\begin{center}
\resizebox{0.5\textwidth}{!}{\includegraphics{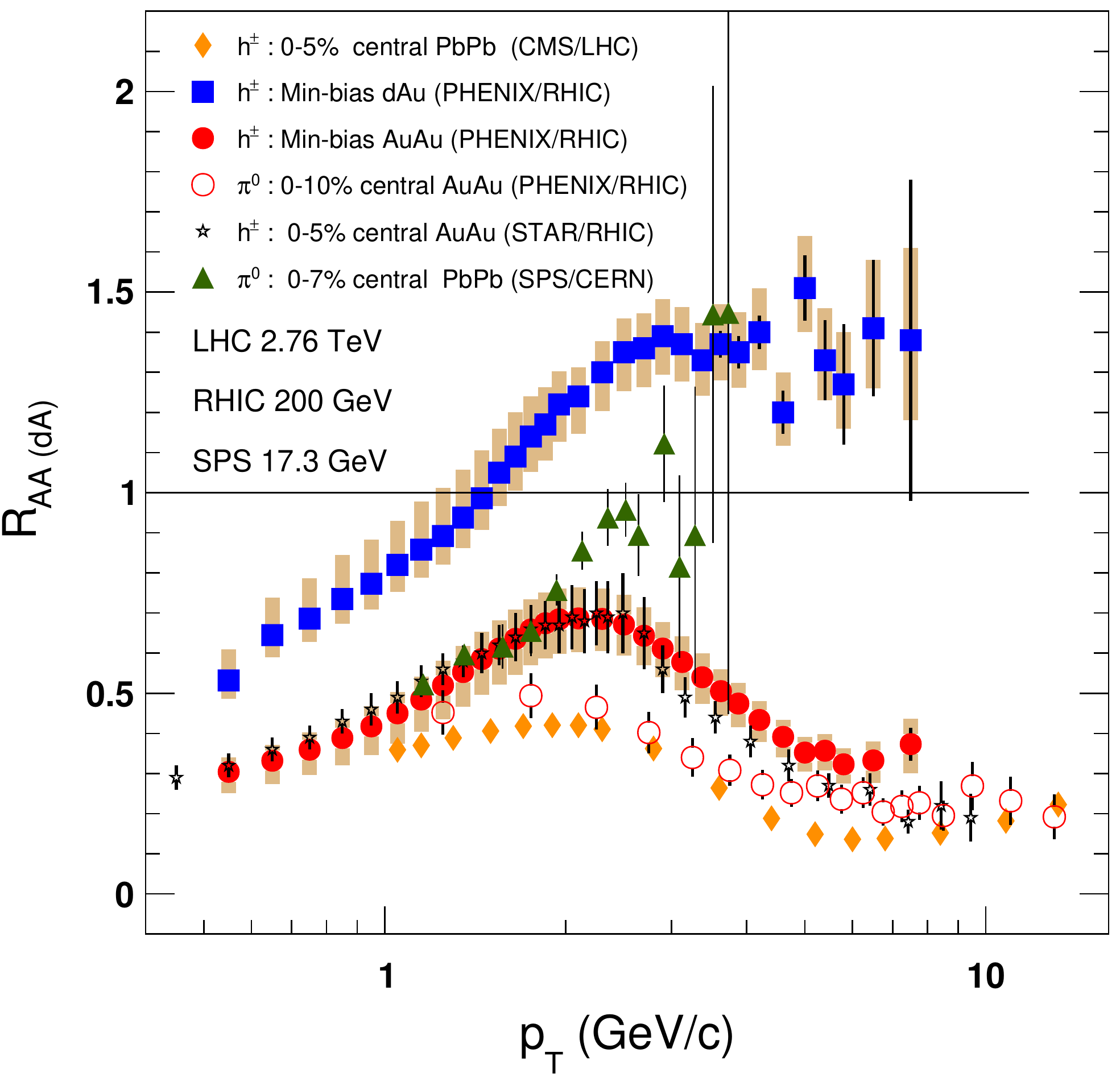}}
\caption{The evolution of the nuclear modification factor with
  center-of-mass energy, from the SPS~\cite{david2004} to RHIC
  \cite{RachidHDR,CompilationRaa_1,CompilationRaa_2,CompilationRaa_3,CompilationRaa_4,RN2011,NN2012} and then to the
  LHC~\cite{CMS}. The error bars correspond to the statistical
  errors. For clarity, the systematic errors are shown as vertical
  bands.}
\label{fig:RaaSPSLHC} 
\end{center}
\end{figure}
}

\def\FigureRaaRHIC{
\begin{figure}
\begin{center}
\resizebox{0.45\textwidth}{!}{\includegraphics{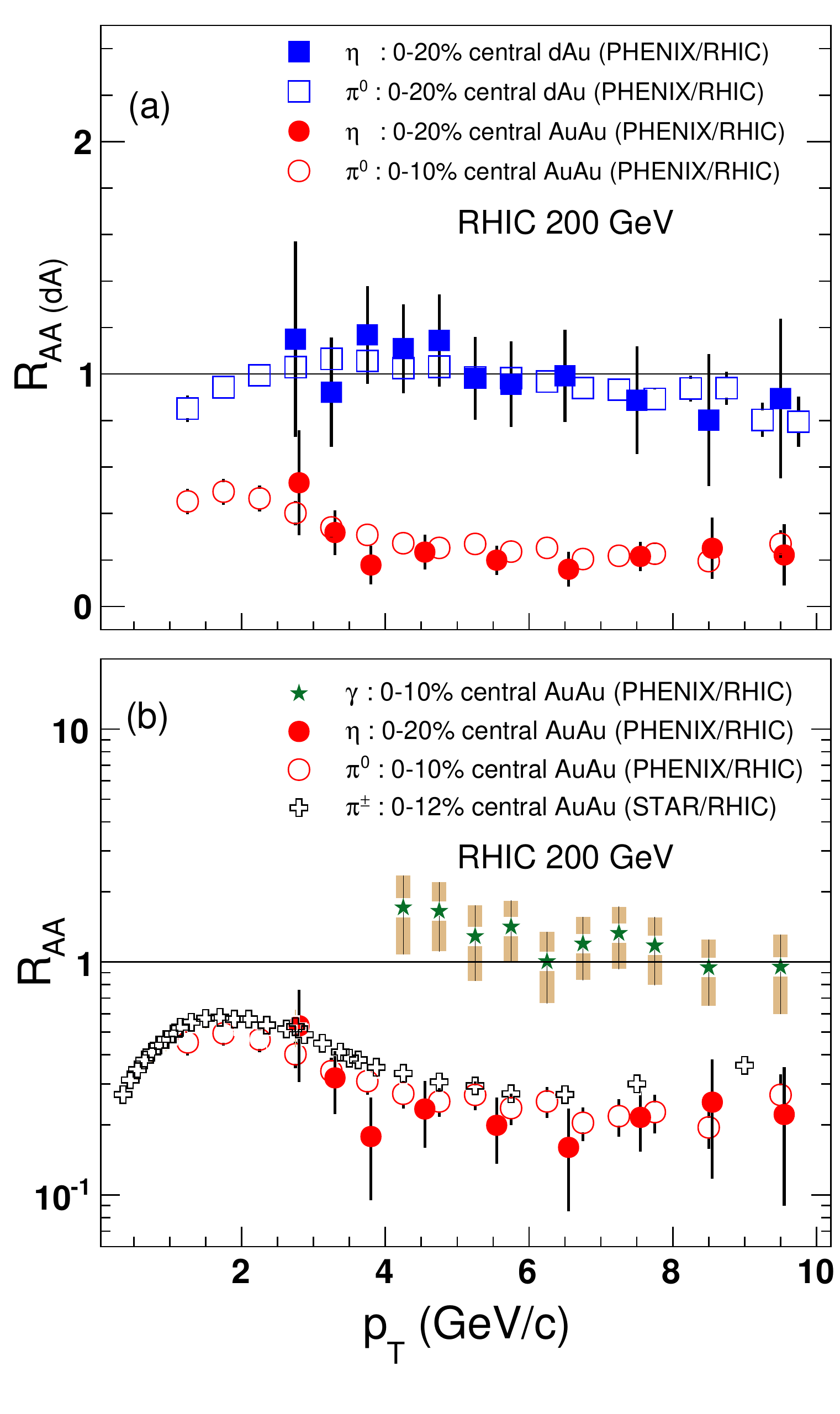}}
\caption{The nuclear modification factor ${\rm R_{AA}}$ from \AuAu
  collisions and ${\rm R_{dA}}$ of \dAu collisions at \snn = 200 GeV
  using Refs. \cite{RachidHDR,CompilationRaa_1,CompilationRaa_2,CompilationRaa_3,CompilationRaa_4,RN2011,NN2012}. Panel (a)
  compares ${\rm R_{AA}}$ with ${\rm R_{dA}}$ for $\pi^{0}$ and $\eta$
  from central \AuAu and \dAu collisions. Panel (b) compares the ${\rm
    R_{AA}}$ of Direct $\gamma$ with ${\rm R_{AA}}$ of $\pi^{0}$,
  $\eta$ and $\pi^{\pm}$ in central \AuAu collisions. The error bars
  correspond to the statistical errors. For clarity, the systematic
  errors are shown as vertical bands.}
\label{fig:RaaRHIC} 
\end{center}
\end{figure}
}

\def\FigureRaaHF{
\begin{figure}
\begin{center}
\resizebox{0.5\textwidth}{!}{\includegraphics{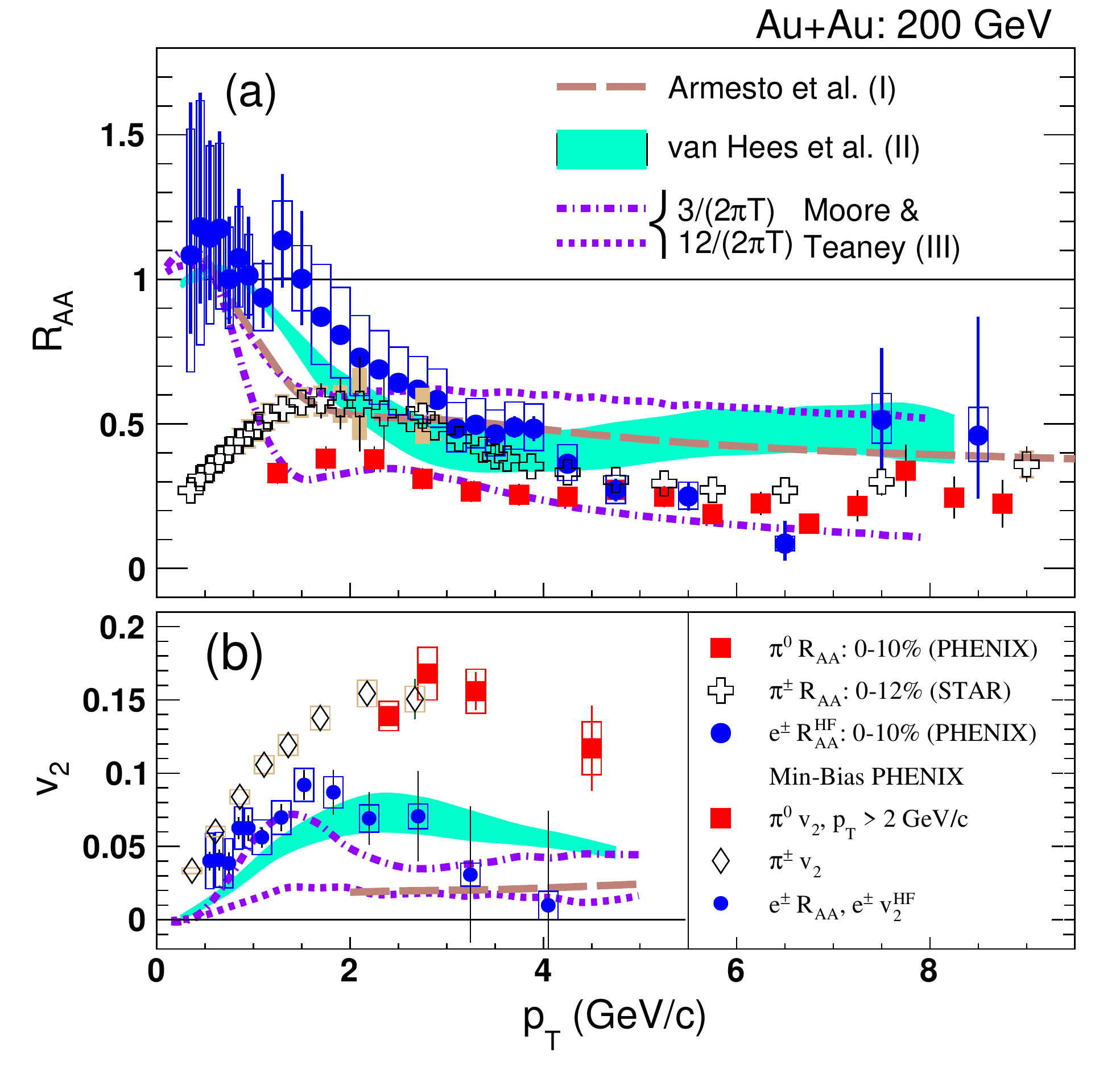}}
\caption{The nuclear modification factor, ${\rm R_{AA}^{HF}}$, for HF
  electrons compared with the ${\rm R_{AA}}$ of $\pi^{0}$ in central
  Au+Au collisions at ${\rm \sqrt{s_{_{NN}}}}$ =\ 200\,GeV, see panel
  (a). Panel (b) considers the anisotropic flow of HF electrons
  v$_{2}^{HF}$ with that of v$_{2}$ of $\pi^{0}$ and $\pi^{\pm}$ in
  minimum-bias Au+Au collisions
  \cite{RachidHDR,CompilationRaa_2,RN2011,NN2012,HeavyPHENIX,HeavySTAR}.}
\label{fig:RaaHF} 
\end{center}
\end{figure}
}

\def\FigureDijets{
\begin{figure}
\begin{center}
\resizebox{0.45\textwidth}{!}{\includegraphics{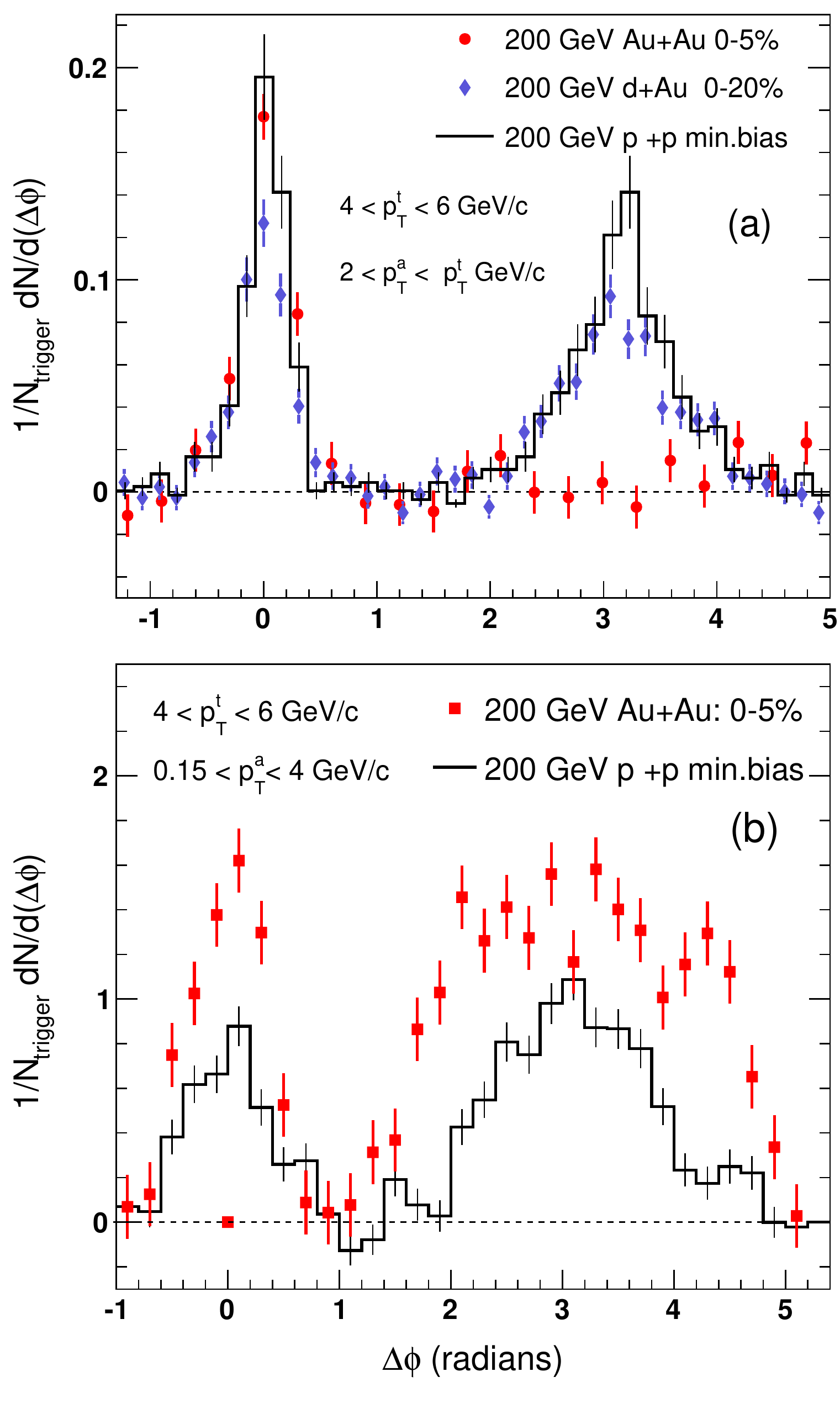}}
\caption{ The two-particle angular correlations measurements in \AuAu,
  \dAu and \pp collisions at \snn = 200 GeV in the presence of a
  trigger particle with \ptt under the condition 4 $<$ \ptt$<$ 6
  GeV/c, and an associated particle with \pta: panel (a) for 2
  $<$\pta$<$\ptt GeV/c, and panel (b) for 0.15 $<$\pta$<$ 4 GeV/c. The
  non-correlated background and the flow background were subtracted
  \cite{RachidHDR,STARdijet_1,STARdijet_2}.}
\label{fig:Dijets} 
\end{center}
\end{figure}
}
\def\TableOne{
\begin{table}[!]
\caption{RHIC operating modes and total integrated luminosity delivered to all experiments for each run \cite{RHICBeam}.}
\begin{center}
\begin{tabular}{|l|l|l|l|l|  }
\hline\hline
\multirow{2}{*}{Run} &\multirow{2}{*}{Species}&\multirow{2}{*}{Total}& 
\multirow{2}{*}{Total} \\
\multirow{2}{*}{} &\multirow{2}{*}{}&\multirow{2}{*}{particle}& 
\multirow{2}{*}{delivered} \\
\multirow{2}{*}{} &\multirow{2}{*}{}&\multirow{2}{*}{energy: \sn}& \multirow{2}{*}{luminosity}\\
\multirow{2}{*}{} &\multirow{2}{*}{}&\multirow{2}{*}{[GeV/nucleon]}& \multirow{2}{*}{[$\mu$b$^{-1}$]}\\
\multirow{2}{*}{} &\multirow{2}{*}{}&\multirow{2}{*}{}& \multirow{2}{*}{}\\
\hline
\multirow{2}{*}
   &   &    &    \\ 
{I (2000)}       & Au+Au    &  56 &  $<$ 0.001  \\ 
                 & Au+Au    & 130 & 20          \\ 
\multirow{3}{*}
                 &   &    &    \\ 
 {II (2001/2002)}& Au+Au    & 200   &  258          \\ 
				 & Au+Au    & 19.6  & 0.4\\                
                 & p+p      & 200   &  1.4$\times$10$^{-6}$ \\ 
\multirow{2}{*}
                 &   &    &    \\
{III (2003)}     & d+Au     & 200 & 73$\times$10$^{-3}$ \\ 
                 & p+p      & 200 & 5.5$\times$10$^{-6}$\\ 
\multirow{3}{*}
				 &   &    &    \\
{IV (2004)}      & Au+Au    & 200   & 3.53 $\times$10$^{-3}$ \\ 
                 & Au+Au    & 62.4  & 67 \\
                 & p+p      & 200   & 7.1$\times$10$^{-6}$ \\ 
\multirow{5}{*}
                 &   &    &    \\
{V (2005)}       & Cu+Cu    & 200   & 42.1$\times$10$^{-3}$ \\ 
                 & Cu+Cu    & 62.4  & 1.5$\times$10$^{-3}$\\ 
                 & Cu+Cu    & 22.4  & 0.02$\times$10$^{-3}$           \\ 
                 & p+p      & 200   & 29.5 $\times$10$^{-6}$\\
                 & p+p      & 410   & 0.1$\times$10$^{-6}$ \\
\multirow{2}{*}
                 &   &    &    \\
{VI (2006)}      &  p+p     &  200.0 & 88.6$\times$10$^{-6}$ \\ 
                 &  p+p     &  62.4  & 1.05$\times$10$^{-6}$\\         
\multirow{2}{*}
                 &   &    &    \\ 
{VII (2007)}     &  Au+Au   &  200.0 & 7.25$\times$10$^{-3}$ \\ 
                 &  Au+Au   &  9.2   &  small \\         
\multirow{3}{*}
                 &   &    &    \\
{VIII (2008)}    &  d+Au    &  200   & 437$\times$10$^{-3}$  \\ 
                 &  p+p    &  200   & 38.4$\times$10$^{-6}$  \\ 
                 &  Au+Au  &   9.2  & small           \\ 
\multirow{3}{*}
                 &   &    &    \\ 
{IX (2009)}      &  p+p     & 500    &  110$\times$10$^{-6}$  \\ 
                 &  p+p     & 200    &  114$\times$10$^{-6}$ \\ 
\multirow{6}{*}
                 &   &    &    \\ 
{X (2010)}       &  Au+Au   & 200     &  10.3$\times$10$^{-3}$   \\ 
                 &  Au+Au   &  62.4   &  544  \\ 
                 &  Au+Au   &  39     &  206  \\ 
                 &  Au+Au   &   7.7   &  4.23  \\ 
                 &  Au+Au   &   11.5  &  7.8    \\ 
\multirow{4}{*}
				 &   &    &    \\ 
{XI (2011)}      & p+p      & 500    &  166$\times$10$^{-6}$  \\ 
                 & Au+Au    & 19.6   &   33.2  \\ 
                 & Au+Au    & 200    &   9.79$\times$10$^{-3}$  \\ 
                 & Au+Au    & 27     &   63.1 \\ 
\multirow{4}{*}
				 &   &    &    \\ 
{XII (2012)}     & p+p      & 200    &   74.0$\times$10$^{-6}$  \\ 
                 & p+p      & 510    &   283$\times$10$^{-6}$  \\ 
                 & U+U      & 193    &   736           \\ 
                 & Cu+Au    & 200    &   27.0$\times$10$^{-3}$   \\           
\multirow{1}{*}
                 &   &    &    \\
{XIII (2013)}    & p+p    & 510    &   1.04$\times$10$^{-9}$  \\ 
                 
\multirow{2}{*}
                 &   &    &    \\
{XIV (2014)}     & Au+Au             & 14.6   &  44.2 \\ 
                 & Au+Au             & 200    &  43.9$\times$10$^{-3}$  \\ 
                 & He+Au             & 200    &  134$\times$10$^{-3}$ \\
\multirow{2}{*}
&   &    &    \\
{XV (2015)}     & p+p           & 200    & 382 $\times$10$^{-6}$  \\ 
                & p+Au          & 200    & 1.27 $\times$10$^{-6}$   \\ 
                & p+Al          & 200    & 3.97 $\times$10$^{-6}$ \\                   
\hline
\hline 
\end{tabular}
\end{center}
\label{tab:rhic}
\end{table}
}
\begin{document}
\title{New State of Nuclear Matter: Nearly Perfect Fluid of
  Quarks and Gluons in Heavy Ion Collisions at RHIC Energies}
\subtitle{From Charged Particle Density to Jet Quenching}
\headnote{Review article} 
\author{R. Nouicer
\thanks{\emph{rachid.nouicer@bnl.gov}}
}
\institute{
Department of Physics, Brookhaven National Laboratory, Upton, NY 11973, United States} 
   
\date{Received: date / Revised version: date}

\abstract{\normalfont\small This article reviews several important
  results from RHIC experiments and discusses their implications. They
  were obtained in a unique environment for studying QCD matter at
  temperatures and densities that exceed the limits wherein hadrons
  can exist as individual entities and raises to prominence the
  quark-gluon degrees of freedom. These findings are supported by
  major experimental observations via measuring of the bulk properties
  of particle production, particle ratios and chemical freeze-out
  conditions, and elliptic flow; followed by hard probe measurements:
  high-\pT hadron suppression, dijet fragment azimuthal correlations,
  and heavy flavor probes. These measurements are presented for
  particles of different species as a function of system sizes,
  collision centrality, and energy carried out in RHIC experiments.
  The results reveal that a dense, strongly-interacting medium is
  created in central \AuAu collisions at \snn = 200 GeV at RHIC. This
  revelation of a new state of nuclear matter has also been observed
  in measurements at the LHC.  Further, the IP-Glasma model coupled
  with viscous hydrodynamic models, which assumes the formation of a
  QGP, reproduces well the experimental flow results from \AuAu at
  \snn = 200 GeV. This implies that the fluctuations in the initial
  geometry state are important and the created medium behaves as a
  nearly perfect liquid of nuclear matter because it has an
  extraordinarily low ratio of shear viscosity to entropy density,
  \mbox{$\eta/s$ $\approx$ 0.12}. However, these discoveries are far
  from being fully understood. Furthermore, recent experimental
  results from RHIC and LHC in small \pA, \dAu and $^{3}$He+Au collision
  systems provide brand new insight into the role of initial and final
  state effects. These have proven to be interesting and more
  surprising than originally anticipated; and could conceivably shed
  new light in our understanding of collective behavior in heavy-ion
  physics. Accordingly, the focus of the experiments at both
  facilities RHIC and the LHC is on detailed exploration of the
  properties of this new state of nuclear matter, the QGP.
} 
\titlerunning{New State of the Nuclear Matter: Nearly Perfect Fluid
  of Quarks and Gluons at RHIC}

\maketitle

\tableofcontents
\section{Physics motivation and RHIC achievements}
At the heart of each atom, there is a nucleus consisting of distinct
nucleons (protons and neutrons). In turn, these nucleons consist of
quarks bound together by the strong interaction, mediated by the
exchange of gluons. Quantum Chromodynamics (QCD) is considered the
fundamental theory for such strong interactions. According to QCD, at
ordinary temperatures or densities this force simply confines the
quarks into composite particles (hadrons) of around
\mbox{10$^{^{-15}}$m = 1 fm} in size (corresponding to the QCD energy
scale \mbox{${\rm \Lambda_{_{QCD}}}$ $\approx$ 200 MeV)}; its effects
are not noticeable at longer distances. However, when the temperature
reaches the QCD energy scale (T of the order of \mbox{10$^{^{12}}$
  Kelvin)} or its density rises to the point where the average
inter-quark separation is less than 1 fm (quark chemical potential
\mbox{${\rm \mu_{q}}$ around 400 MeV)}, hadronic matter under
extremely dense and hot conditions undergoes a phase transition to
form a Quark Gluon Plasma (QGP) in which quarks and gluons no longer
are confined to the size of a hadron \cite{Shuryak801,Shuryak802}. Knowing the
exact nature of quark confinement in hadrons is crucial, and yet this
remains a poorly understood aspect of the quark-gluon description of
matter. The QGP is believed to have existed during the first
microseconds after the Big Bang~\cite{EarlyUniverse}, and an
understanding of its properties could provide valuable insights on the
evolution of our Universe. Figure~\ref{fig:Phase} presents the phase
diagram of QCD matter, showing the possible states adopted by quarks
and gluons \cite{phase}.  
\FigurePhase

Understanding the properties of strongly interacting matter at high
temperatures has been a central goal of many researchers' numerical
simulations of lattice QCD. Figure~\ref{fig:QCD} presents recent
results from lattice QCD calculations, wherein the energy density is
normalized by temperature \mbox{(${\rm T^{4}}$)} as a function of
temperature for a system of $n_{\rm f} = $ 2 + 1 flavors of dynamical
quarks \cite{Sza2010,Sza2014,HotQCD}. These calculations reveal a
rapid increase in the number of degrees of freedom associated with
this deconfinement of quarks and gluons from the hadronic chains. The
transition point is at a temperature ${\rm T_{c} = 154 \pm 9}$ MeV and
energy density ${\rm \epsilon}_{c} =$~0.18~to~0.5~${\rm GeV/fm^{3}}$
\cite{HotQCD}. At higher temperatures above the transition, the energy
density is still seen to fall about 14\% below the ideal
Stefan-Boltzmann value for a non-interacting quark/gluon gas (as shown
by the arrow), indicating some strong self-interaction remaining at
these temperatures. Therefore, scientists have experimentally searched
for the signatures both of QGP formation and the in-medium effects of
hadron properties. It was proposed that the required high densities
could be achieved via relativistic heavy-ion collisions
\cite{Greiner1975}.  Over the last decades, relativistic heavy-ion
collisions have been studied experimentally at increasingly high
center-of-mass energies at the Brookhaven Alternating Gradient
Synchrotron (AGS, \snn $<$ 5 GeV), the CERN Super Proton Synchrotron
(SPS, \snn $\le$\ 20 GeV), and presently, at the Brookhaven
Relativistic Heavy Ion Collider (RHIC, \snn $\le$ 200 GeV) and the
Large Hadron Collider (LHC at CERN, \snn = 2.76 TeV). Simple
estimations of the initial energy densities expected for RHIC
collisions \cite{Bjorken} are many times higher than the threshold for
QGP formation from the aforementioned lattice estimations.  
\FigureQCD

For instance, under the RHIC project, an accelerator was constructed
at Brookhaven National Laboratory (BNL) from 1991 to 1999. RHIC was
designed as a heavy-ion machine, able to provide collisions over a
large range of energies. We have attained collisions of gold-gold
(\AuAu), copper-copper (\CuCu), uranium-uranium (\UU), copper-gold
(\CuAu), deuteron-gold (\dAu), helium-gold (\HeAu), proton-gold (\pAu)
and proton-Aluminum (\pAl) at center of mass energies from \snn~= 7.7
to 200~GeV. Furthermore, a polarized capability was added to RHIC
allowing transverse and longitudinal polarized protons to collide at
energies from \snn = 62.4 to 510~GeV. Researchers at RHIC have made a
major physics discovery, namely the creation of a new form of matter
in high-energy central gold--gold collisions. This matter, dense and
strongly interacting, is called the strongly coupled quark-gluon
plasma, or sQGP.  This finding of a new form of matter was published
in four independent white papers \cite{RHIC_1,RHIC_2,RHIC_3,RHIC_4}
from the four RHIC experiments: BRAHMS (Broad RAnge Hadron Magnetic
Spectrometers), PHENIX (Pioneering High Energy Nuclear Interaction
eXperiment), PHOBOS (not an acronym) , and STAR (Solenoidal Tracker At
RHIC). RHIC's accelerator also met and exceeded its specifications;
namely, it attained its energy goals and exceeded both the heavy-ion
\cite{AuAuDesign,RHICBeam} and polarized proton
\cite{RHICBeam,ppDesign} luminosity.  The former luminosity is more
than a factor of 25 \cite{AuAuDesign,RHICBeam} greater than that
originally designed.  Table~\ref{tab:rhic} summarizes the achievements
of RHIC accelerator over the years \cite{RHICBeam}.  Without a doubt,
the physics program at RHIC has enabled remarkable advances in the
study of hot strongly interacting matter
\cite{RHIC_1,RHIC_2,RHIC_3,RHIC_4,McGyl2005}.  The extended particle
momentum range at RHIC allowed the use of hard probes to study the
behavior of the created medium that was difficult to access at lower
collision energies. The hard penetrating probes now are one of the
major experimental tools successfully exploited by RHIC and LHC
Collaborations \cite{RHIC_1,RHIC_4,LHCATLAS2014,LHCALICE2014}.  It
should also be pointed out that, recently, new physics results have
emerged due to the availability of higher colliding beam energies at
the LHC.  For example, the studies of two-particle azimuthal
correlation functions in the highest-multiplicity \pp collisions at
the LHC show an enhancement of particle pair production at relative
azimuth $\Delta\phi~\simeq$~0, producing a ``ridge'' structure at the
near-side \cite{Kh2010}.  However, in peripheral \pPb collisions, in
which only a few nucleons take part and are nominally similar to \pp
collisions, a ridge structure is observed on the away-side
($\Delta\phi~\simeq$~$\pi$) \cite{Aad2013}.  The different results
between \pp and peripheral \pPb collisions are possibly caused by
contributions from nuclear effects. In central \pPb collisions, a
double ridge structure is observed, which is consistent with \PbPb
collisions \cite{Mil2014_1,Mil2014_2}.  Another result in the LHC
energy regime, \PbPb collisions at \snn = 2.76 TeV, is the nuclear
modification factor \RAA of inclusive \Jpsi production measured in the
dielectron channel as a function of centrality
\cite{LHCALICE2014}. These results are compared with the inclusive
\Jpsi \RAA measured at mid-rapidity by PHENIX in \AuAu collisions at
\snn = 200 GeV \cite{Adare2007}. The inclusive \Jpsi production in
\PbPb collisions show less suppression compared to results at RHIC
energies, and this behavior is in qualitative agreement with a
regeneration scenario at LHC
energies~\cite{Regeneration_1,Regeneration_2}.  \TableOne

In this article, we highlight the RHIC results underlying the major
experimental observations from the heavy-ion program, i.e., the
identification of a new form of nuclear matter.  This required the
acquisition of novel and uniquely different measurements from those
seen before in heavy-ion collisions at lower energies. To set the
foundation, we have focused this review paper on measurements for
different particle species as a function of: 1) system size, 2)
collision centrality, and 3) energy, all carried out in RHIC
experiments.  As RHIC is currently still operational, some data
presented here has not been available to authors of past review
articles.  As such the importance of this common set of observables,
with a clear interpretation across the broad system size and beam
energies, presented here may not have been emphasized. These findings
are supported by major empirical observations and elucidated in the
present work from both soft and hard probe sectors. For soft probes,
measurements on the production of bulk particles and the initial
conditions are presented, followed by particle ratios and chemical
freeze-out conditions. These measurements are very important as they
support the characterization of the system produced in such collisions
and provide basic constraints to theoretical models. To complete the
soft-sector discussion, we present measurements of collective flow
that are related indirectly to the thermodynamical or hydrodynamical
properties of the new form of nuclear matter, a nearly perfect
fluid. This discussion is then followed by hard probe measurements:
high-\pT\ hadron suppression and dijet fragment azimuthal correlations
which lead to the observation of the jet quenching in relativistic
heavy-ion collisions viz., one of the most remarkable discoveries at
RHIC. These measurements of hard probes afford experimental evidence
of highly interacting dense matter created in central \AuAu collisions
at \snn = 200 GeV. Interpretation of the modification of bound states
of heavy quarks, charmonium and bottomonium, is heavily impacted by
many competing effects in heavy ion collisions including large initial
state nuclear effects not directly related to the formation of the
quark-gluon plasma and a full understanding of these observables has
not yet been formulated
\cite{Vogt2005,Khar1997,Eskola2009,Nagle2011}. This review of hard
sector observables thus focuses on quenching observables which are
more simply related to the properties of the quark-gluon
plasma. Global properties of charged hadron production including the
empirical evidence of QGP a nearly perfect fluid are presented in
Section~\ref{sec:Global}. Experimental observations of the creation of
dense medium, and its opacity, based on high-\pt hadron suppression,
and dijet fragment azimuthal correlations are elucidated in details in
Section~\ref{sec:dense}.
\section{Global properties of charged hadron production \label{sec:Global}}
The heavy-ion collisions at the highest RHIC energy, i.e. \AuAu at
\snn = 200 GeV, heralded a new era of opportunities for studying
hadronic matter under conditions of high energy density.  The theory
of strong interactions, QCD, can be used to compute processes
perturbatively in the limit of short distances. At present, however,
we cannot compute, from first principles, many interesting quantities
at large distance. Therefore, it is important to characterize the
collisions using global extrinsic observables. Measurements of charged
hadron multiplicity and transverse energy distributions and collective
flow in heavy-ion collisions afford us information on the initial
energy density, initial collision geometry and the entropy production
during the system's evolution. All are sensitive to a variety of
physics processes responsible for multi-particle production. This
knowledge is important for constraining model predictions and
indispensable for understanding and estimating the accuracy of the
more detailed measurements, for example, those of jets or quarkonia
production.
\FigureSoftpeak

Figure~\ref{fig:Softpeak} shows the measured transverse momentum, \pT,
distribution of charged hadrons produced in the 0-15\% most central
\AuAu collisions at \snn = 200 GeV~\cite{Rachid2003}. The distribution
illustrates a ``bulk'' and ``tail'' respectively often associated to
``soft'' and ``hard'' parton-parton scattering.  However, there is no
clear separation between ``soft'' and ``hard'' processes. For this
reason, we require an analysis undertaken as a function of centrality,
transverse momentum, energy, and system sizes to better understand
particle production. We note that most of the particles under
investigation correspond to ``thermal" pions (\pT~$<$~2~GeV/{\it c})
and, in general, such thermal hadrons make up about 95\% of the
observed particle multiplicity referred to as the bulk of hadron
production. Their distribution in phase space of pseudorapidity
($\eta$) as a function of collision centrality and energy is presented
in Subsection~\ref{sec:density}. These measurements led to the first
insights into the overall reaction dynamics, and also set the stage
for considering more rare signals, embedded in this thermal bulk
production.  
\FigureGlauber
\subsection{Charged particle density distributions and initial conditions \label{sec:density}}
One important observable in heavy-ion interactions is the charged
particle pseudorapidity density, \dNch. The pseudorapidity, ${\rm \eta
}$, is defined as \mbox{${\rm \eta = -ln[tan(\theta/2)]}$}, where
\mbox{${\rm \theta }$} is the emission angle relative to the direction
of the beam. This quantity (\dNch) is proportional to the entropy
density at freeze-out. Since the entropy density of a closed system
will be non-decreasing \cite{Ahmad2013}, the pseudorapidity density
provides information on the partons' initial-state density and any
further entropy produced during subsequent evolution
\cite{RachidMoriond2002}. In describing the collision of two nuclei,
several variables are used to quantify the collision's centrality
including, the number of participants nucleons (number of interacting
nucleons), \Np, and the number of binary collisions (number of
nucleon-nucleon collisions), \Nc. Both are given in terms of the
impact parameter, \bim (see Figs. \ref{fig:Glauber} and
\ref{fig:VSch}(a)), the distance between the centers of two colliding
nuclei. A small value of $b$ corresponds to more central collisions,
whilst a large value to peripheral collisions. The impact parameter is
not directly measurable but experiments at BNL (AGS, RHIC) and CERN
(SPS, LHC) showed that the relation between number of charged
particles produced (and the energy of the produced particles in the
transverse direction, \ET) and on average the number of nucleon
participants, \avgNp, is monotonic. Thus, since the dependence of \Np
on the impact parameter can be calculated with some precision for a
given nuclear density distribution, we can get a handle on the
collision's centrality. The Glauber model is a semi-classical model
picturing nucleons with fixed transverse positions moving in the
collision direction in a straight path
\cite{Glauber1970,Shukla2003}. It provides a geometrical description
of multiple nucleon collisions, and is used to relate the impact
parameter to the number of participants and the number of N+N
collisions, based on the assumption of constant inelastic cross
section for each subsequent collision. Figure~\ref{fig:Glauber} shows
the minimum-bias multiplicity (\Nch) distribution used for selecting
collision centrality. The minimum-bias yield was cut into successive
intervals, starting from the maximum value of \Nch. The first 5\% of
the high \Nch events correspond to the top 5\% central collisions. The
relationship of centrality and impact parameter with the number of
participating nucleons was elaborated in Glauber-type Monte Carlo
calculations employing the Wood-Saxon nuclear density distributions
\cite{Mich2007,Shou2015}.
\FigureMultMid

Figure~\ref{fig:MultMid}(a) shows the measured densities of charged
particle near the mid-rapidity region, \dnchmid, as a function of
center--of--mass collision energy (\snn) for \AuAu (RHIC: \snn = 19.6,
56, 62.4, 130 and 200 GeV), and \CuCu (RHIC: \snn = 22.4, 62.4 and 200
GeV)
\cite{RachidMoriond2002,RachidHDR,AuAufrag,Panic2006,CuCu2008}. The
measurements present the 0-6\% most central collisions at mid-rapidity
region and \avgNp\ is the average number of participant nucleon pairs,
for \AuAu and \CuCu.  The results (see Fig.~\ref{fig:MultMid}(a))
suggest that the particle density rises approximately logarithmically
with energy. Comparing the findings for \AuAu and \CuCu indicates
that, for the most central events in symmetric nucleus-nucleus
collisions, the particle density per nucleon participant pair does not
depend on the size of the two colliding nuclei but only on the energy
of the collision \cite{Panic2006,CuCu2008}.  For comparison, the RHIC
data is compared to those obtained at the CERN SPS for \PbPb
collisions at \snn = 12.3 and 17.3
GeV~\cite{SPSdNdEta1,SPSdNdEta2}. We observe that at maximum RHIC
energy, \snn = 200 GeV, is a 98\% higher particle density per
participant pair in near $\eta$~=~0 than at the maximum SPS energy,
17.3 GeV. General arguments based on Bjorken’s estimate~\cite{Bjorken}
suggest that this increase should correspond to a similar increase in
the maximal energy density achieved in the collision.  \FigureMult4pi

Based on the \AuAu data in Fig.~\ref{fig:MultMid}(a) obtained at RHIC
energies, one can establish a power law fit to predict (extrapolate)
the results of \dNch\ at the mid-rapidity region \mbox{($|\eta|<$ 1)}
to \PbPb central collisions at the LHC's energy 5.5 TeV. The power law
fit to the RHIC data is shown as the solid curve in
Fig.~\ref{fig:MultMid}(a) and \ref{fig:MultMid}(b).\\ 
The corresponding of the power law fit function is:
\begin{equation}
{\rm f^{Pow}_{AA} = 0.78 \times (\sqrt{s_{_{NN}}})^{0.3}.} 
\end{equation}
The \snn is in GeV unit. We observe that the power law fit has a good
agreement with the SPS and RHIC data, allowing one to extrapolate the
scaled density per nucleon participant pair for 0-6\% central \PbPb
collisions at LHC energy, 5.5~TeV, in the mid-rapidity region
($|\eta|<1$):
\begin{eqnarray*}
{\rm f^{Pow}_{AA} (5.5\ TeV) } &=&  {{\rm \frac{1}{\langle N_{part}/2\rangle}\frac{dN_{ch}} {d\eta}}  
(\rm Pb+Pb\ at\ 5.5\ TeV)}\\
{\rm f^{Pow}_{AA} (5.5\ TeV) } &=&  10.3 \pm 0.26
\end{eqnarray*}
where the uncertainties are propagated from the power law fit
parameters.  Using a Glauber model calculation for the 0-6\% most
central \PbPb collisions at 5.5 TeV (the total inelastic cross section
used in the Glauber model calculation is \mbox{$\sigma_{_{NN}}$~= 72
  mb}), a value of \mbox{\avgNp = 381 $\pm$ 11} is obtained, from which the
unscaled charged particle pseudorapidity density, i.e. using the
power law extrapolation, at mid-rapidity ($|\eta|<1$) can be deduced:
$${{\rm \frac{dN_{ch}} {d\eta} (6\%\ central, Pb+Pb\ at\ 5.5\ TeV) =
    1962\ \pm 50}}$$ The results from ALICE (A Large Ion Collider
Experiment) at LHC ~\cite{ALICEPb276} on charged particles density at
mid-rapidity region and for 0-5\% most central \PbPb collisions at
\snn = 2.76 TeV corresponds to the following:
\begin{equation*}
{{\rm \frac{1}{\langle N_{part}/2\rangle} \times \frac{dN_{ch}}
    {d\eta}} (\rm Pb+Pb\ at\ 2.76\ TeV) = 8.4 \pm 0.3}
\end{equation*}
As shown in the Fig.~\ref{fig:MultMid}(b), the power law fit function
and the ALICE/LHC results for central \PbPb at 2.76 TeV agree with
this extrapolated trend.

The calculations of the Impact Parameter dipole Saturation model
(IP-Sat model) \cite{IPCGC2012} for charged particle density,
\dnchmid, as a function of energy from RHIC to LHC are represented in
Fig.~\ref{fig:MultMid}(b) by the dashed curve. We observe that the
IP-Sat model is in reasonably good agreement with the data
\cite{IPCGC2012}.  In these theoretical calculations (IP-Sat model),
\dNch\ is estimated to be approximately 2/3 times the initial gluon
density, ${\rm dN_{g}}$/dy.  Further, this indicates that in IP-Sat
model the saturation scale grows roughly as \mbox{$\langle Q_S^2
  \rangle \sim (\sqrt{s})^{0.3}$} if a fixed value of
\mbox{$\alpha_s=0.2$} is assumed as in Ref. \cite{IPCGC2012}.  In the
Color Glass Condensate (CGC) framework, the produced gluon
multiplicity ${\rm dN_{g}}$/dy $\sim$ $\langle Q_S^2
S_\perp\rangle/\alpha_s$ is expected where $Q_S^2$ is the minimum
saturation scale of the two colliding nuclei, and $S_\perp$ is the
overlap area \cite{Dima2001,Kov2015}.  \FiguredNdEtaCuCuAuAu

Figure~\ref{fig:Mult4pi} shows the measured \dNch of primary charged
particles over a broad range of pseudorapidity, ${\rm |\eta|<}$~5.4,
for \AuAu \cite{RachidHDR,RachidMoriond2002,AuAufrag} and \CuCu
\cite{Panic2006,CuCu2008} collisions under a variety of collision
centralities and RHIC energies. The data from \dAu
\cite{RachidQM2004,dAuPRL,dAuPRC} and \pp\ \cite{RachidQM2004,PHOBIG}
at RHIC energies also are shown. Both the height and width of the
\dNch distributions increase as a function of energy as observed in
the \AuAu and \CuCu collisions. The \AuAu, \CuCu, \dAu and \pp\ data
at all energies were obtained with the same detector setup in the
PHOBOS experiment at RHIC \cite{RachidNIM,PHONIM}. This situation is
optimal as common systematic errors cancel each other out in the
ratio. With this configuration, we were able to examine
comprehensively both particle production in \CuCu and the \AuAu
collisions for the same \Np, for the same fraction of total
inelastic cross sections, and for the same geometry in both systems.
\FigureRatioBrahms

Figure~\ref{fig:dNdEtaCuCuAuAu} presents a comparison of \dNch
distributions for \AuAu and \CuCu collisions at similar energies, \snn
= 22.4 (19.6), 64.2 and 200 GeV. The \dnch distributions are
normalized to \Np. The goal from the measurements presented in
Fig.~\ref{fig:dNdEtaCuCuAuAu} is to study the sensitivity of the full
shape of the pseudorapidity distributions in \AuAu and \CuCu
collisions when the overlap collision regions in both systems are
selected: (1) for the same number of nucleon participants (\Np), and
(2) for the same fraction (\Np/2A), which should reflect the volume of
overlap region as a fraction of the total nuclear volume where A is
the mass number of the colliding nuclei. The first comparison is made
for centrality bins chosen such that the average number of
participants in both systems is the same,
Fig~\ref{fig:dNdEtaCuCuAuAu}(a)-(c). This comparison reveals that, at
200~GeV, the height and the width of the \dNch distributions in both
systems agree within systematic errors, and that at 62.4 and 22.4 GeV,
the distributions agree within systematic errors at mid-rapidity but
not in the fragmentation regions (i.e., high ${\rm |\eta|}$). The
\dNch distributions of \AuAu collisions at 19.6 and 62.4 GeV were
interpolated linearly in \lnsnn\ to obtain the scaled \AuAu data at
22.4 GeV. We observed that the production of charged particles in the
high ${\rm |\eta|}$ region is increased in \AuAu compared to that in
\CuCu collisions.  This increase may be attributed to the two excited
nuclear remnants being bigger in the former collisions relative to the
latter.  This effect is most visible at the lowest energies where the
broad $\eta$ coverage gives access to $|\eta| > y_{beam}$. The second
comparison, \mbox{Fig.~\ref{fig:dNdEtaCuCuAuAu}(d)-(f)}, shows \dNch
distributions in centrality bins with similar values of \Np/2A (and
automatically have matching \Nsp/2A\ values, where \mbox{\Nsp = 2A -
  \Np} is the number of non-participating nucleons, the spectators) in
\AuAu and \CuCu systems. We observe a better matching
over the full $\eta$ coverage and for the three energies, 22.4, 62.4,
and 200 GeV, see \mbox{Fig.~\ref{fig:dNdEtaCuCuAuAu}(d)-(f)}, with the
results shown for central collisions in both systems. Similar
comparisons (same \Np/2A) for more peripheral bins lead to the same
conclusion as shown in
\mbox{Fig.~\ref{fig:dNdEtaCuCuAuAu}(g)-(i)}. The comparison presented
in the Fig.~\ref{fig:dNdEtaCuCuAuAu} reveals an interesting feature:
the \dnchNp\ distributions of central \AuAu and central \CuCu
collisions (as well peripheral \AuAu and peripheral \CuCu collisions)
are similar over the detector's full coverage (${\rm |\eta|~<}$~5.4),
and for the three energies, 22.4, 62.4 and 200 GeV, implying that the
production of charged particles, normalized to the number of nucleon
participants, is mostly driven by the volume of overlap region, \Np/2A
(i.e., the fraction of the total nuclear volume that interacts).

One question is whether the mechanisms in the limiting fragmentation
region (i.e. high ${\rm |\eta|}$) are distinct from those at
mid-rapidity region (${\rm |\eta| < 1}$). From the results in the
Figs.~\ref{fig:Mult4pi} and \ref{fig:dNdEtaCuCuAuAu} for the \AuAu,
\CuCu, \dAu, and \pp\ collisions, there is no obvious evidence for two
separate regions at any of the RHIC energies.  However, to date no
such anomalies have been noted at the energies of the AGS, SPS, RHIC,
or LHC. So far, all results on charged particle pseudorapidity
densities point to a rather smooth evolution in centrality,
peseudorapidity, and \sn. It had been imagined, that a phase
transition would manifest as a non-monotonic behavior in these
observables; such sharp features are absent, but that alone does not
rule out the presence of a phase transition~\cite{Thomas}.  It should
be noted that mechanisms of charged particle production for \dNch
(Fig.~\ref{fig:Mult4pi}) are mostly driven by ``soft'' processes
(Fig.~\ref{fig:Softpeak}). Also note that pseudorapidity distributions
correspond to inclusive measurements of charged hadrons. It would be
interesting to have the same measurements for identified particles
over a large coverage (i.e. ${\rm |\eta| < 5.4}$) as the behavior of
particles species in the two regions, mid-rapidity and fragmentation,
may differ.  
\FigureRatiotherm
\FigureRatioStar
\subsection{Particle yields and chemical freeze-out conditions}
Chemical freeze-out is the stage in the evolution of the hadronic
system at which inelastic collisions cease and the relative particle
ratios become fixed. In a chemical analysis within the grand-canonical
ensemble, the statistical-thermal model with conservation laws
requires at least five parameters as input: volume V, the chemical
freeze-out temperature \Tch, baryon-, strangeness- and charge chemical
potentials \mb, \ms and \mq respectively. These parameters determine
the particle composition in the hadronic final state.  After chemical
freeze-out, the particle composition inside the fireball is fixed, but
elastic collisions keep the system interacting until the final,
thermal (kinetic) freeze-out. At this stage the produced particle
spectra are fixed (modulo weak decays) and carry information about the
phase-space distribution in the final state of the fireball.
Therefore, the transverse momentum spectra determine the parameters of
the thermal freeze-out. Thereafter, statistical interpretation of
particle production becomes an appropriate approach for evaluating
heavy-ion collisions at high energies because of the large
multiplicities of hadrons which are created. One can assume that the
nuclear matter created in these collisions form an ideal gas that can
be characterized by a grand-canonical ensemble. Using thermodynamic
concepts to describe multi-particle production has a long
history~\cite{Hagedorn1965}. The concept of a temperature applies only
to systems in at least local thermal equilibrium. The assumption of a
locally thermalized source in chemical equilibrium can be tested by
applying statistical thermal models to describe the ratios of various
emitted particles.

The BRAHMS collaboration has reported results on particle ratios as a
function of rapidity~\cite{BRAHMSRatio} as shown in
Fig.~\ref{fig:RatioBrahms} which depicts the rapidity dependence of
the ratios $\bar{p}/p$, $\pi^{-}/\pi^{+}$ and $K^{-}/K^{+}$ produced
in \AuAu collisions at the maximum RHIC energy, \snn = 200 GeV.  The
ratios were obtained by integrating over both \pT\ and centrality.  A
strong rapidity dependence of the $\bar{p}/p$ ratio is evident,
dropping from 0.75 $\pm$ 0.06 at the mid-rapidity region ($y =0$) to
0.23 $\pm$ 0.03 at $y \simeq$ 3. This is a large deviation from a
boost invariant source where the ratios are constant; this indicates
that as we move away from the low net-baryon central region, a baryon
rich fragmentation region is reached. The $\pi^{-}/\pi^{+}$ ratio is
consistent with unity (flat) over the considered rapidity range, while
the $K^{-}/K^{+}$ ratio drops almost by 30\% at y $=$ 3 from its
mid-rapidity value. The $\bar{p}/p$ and $K^{-}/K^{+}$ ratios are
essentially constant in the rapidity interval $ y = [0,1]$.

As mentioned above, the measured set of particle ratios (or yields) at
mid-rapidity lends itself to an analysis in terms of a model based on
assuming the system is in chemical and thermal equilibrium.
Figure~\ref{fig:Ratiotherm} presents a recent comparison of the RHIC's
experimental results and statistical thermal model calculations of
hadron yields produced in \AuAu collisions at \snn = 200 GeV
\cite{Anton2013}. The measurements were taken in the mid-rapidity
region $|\eta| < 1$. They demonstrate quantitatively the high degree
of equilibration achieved for hadron production in central Au$+$Au
collisions at maximum RHIC energy. The numerical agreement between
these calculations and the present measurements is excellent.  The
statistical thermal model reveals that the apparent species
equilibrium was fixed at a temperature \Tch = 162 MeV and \mb = 24
MeV. The ratios involving multi-strange baryons are well reproduced,
as is the $\phi$ yield. Even the relatively wide resonances such as
the K$^{*}$'s fit well with the picture of a chemical freeze-out. The
small value of the chemical potential, \mb, indicates a small net
baryon density at mid-rapidity at the RHIC which was confirmed by
measurements of net-baryon rapidity distribution in \AuAu at \snn =
200 GeV \cite{BRAHMSStopping}.  
\FigureRatioLEP

Figure~\ref{fig:RatioStar} shows the STAR collaboration thermal model
fit of temperature and baryon potential at chemical freeze-out as a
function of collision centrality (\dnch) in collision systems \AuAu,
\CuCu and \pp \cite{StarPRC2011}. The chemical freeze-out temperature,
\Tch, is around 155 MeV and seemingly shows no dependence on the
centrality and also no sensitivity to the colliding systems, yielding
the same results for \AuAu and \CuCu data. The baryon chemical
potential, \mb, also showed no difference between the fits to \AuAu
and \CuCu data.  To study the validity of the statistical thermal fit
considering a Grand Canonical formulation, the same fit was applied to
the \pp\ data. The chemical freeze-out temperature that results from
this fit is slightly lower, \mbox{\Tch $\simeq$ 150~MeV}
\cite{StarPRC2011}.

This equilibrium population of species occurs both in elementary and
nuclear collisions \cite{Stock2006,Munz2004}. It was shown in \ee
annihilation data at \s = 91.2~GeV LEP energy, shown in the
Fig.~\ref{fig:RatioLEP}, are reproduced by the statistical
hadronization model in its canonical form \cite{Klim2010,Beca2002} and
the derived temperature was \Tch = 165~MeV, in agreement with the
limiting temperature predicted by Hagedorn \cite{Hagedorn1965} to
occur in any multi-hadronic equilibrium system once the energy density
approaches 0.6 GeV/fm$^3$. Thus, the upper limit of hadronic
equilibrium density corresponds closely to the upper limit of energy
density, $\epsilon_{\rm crit} =$ 0.18 to 0.5~GeV/fm$^3$ of partonic
equilibrium matter, according to lattice QCD \cite{HotQCD,Kars200}.
This universal value of \Tch~= 150 to 170 MeV is remarkably close to
the critical temperature for the quark-hadron transition from Lattice
QCD and it appears to be the universal hadronization temperature for
\ee, \pp and heavy-ion collisions at high
energy~\cite{Klim2010,Ulri2006}. To what extent this hints at the
thermal behavior in \ee and \pp collisions is an important question,
which is still under debate in our field. More recent data in \dAu and
\pPb as well in \pp at RHIC and LHC energies provide new insight into
the study of thermalization behavior in smaller systems compared to
\AuAu and \PbPb.  
\subsection{QGP a nearly perfect fluid}
The establishment of a new form of nuclear matter demands the
observation of new collective properties distinct from previous
measurements. The flow pattern of the thousands of particles produced
in heavy-ion collisions is a key observable being adapted to search
for new collective aspects‎
\cite{Greiner1975,Stok79,Stok82,Stok86,Reis77}. The properties of
collective flow examine two of the conditions required for satisfying
the validity of the QGP hypothesis. The first is the degree of
thermalization. Until now, no results are available from Lattice QCD
regarding the non-equilibrium dynamics of a QGP.  Nonetheless‎, the
evolution of bulk matter from some initial conditions could be
calculated through the equations of viscous relativistic hydrodynamics
provided that local equilibrium is preserved. These equations can be
approximated further by perfect fluid equations when the corrections
due to viscosity could be neglected. This is conceivable when the
scattering mean free paths are small compared to the scale of the
spatial gradients of the fluid.  The second condition is the validity
of the numerically determined equation of state, i.e. the relationship
between energy density and pressure. Under a specific initial boundary
condition, the future evolvement of the matter then can be predicted
\cite{McGyl2005}.  To a good approximation, the data on elliptic flow
confirms the idea that local thermal equilibrium is reached at RHIC
energy and that the flow pattern is completely in agreement with
numerical determinations of the equation of state from QCD.
\FigureVSch \FigureVEta

However, collisions of two nuclei at relativistic energies create a
partonic medium at extreme temperatures and densities, which will cool
down by expansion and undergo a phase transition from a QGP back to
hadronic matter. Studying collective behavior (like measurements of
anisotropic particle flow) of strongly interacting matter in such
conditions recently became one of the most active and fascinating
topics in physics.  The measurements of anisotropic particle flow and
jet quenching at RHIC revealed a deconfined state of matter at high
temperature and partonic density, that is characterized best as a near
perfect fluid, i.e. a collective state with an extremely low ratio of
shear viscosity to entropy density
\cite{RHIC_1,RHIC_2,RHIC_3,RHIC_4}.
\subsubsection{Anisotropic particle flow}
The measurements of anisotropic particle flow, being connected to the
collective behavior of the system, is an important observable in
relativistic heavy-ion collisions as it signals the presence of
multiple interactions between the constituents of the created
matter. It was studied extensively in nucleus-nucleus collisions at
the SPS and RHIC as a function of pseudorapidity, centrality,
transverse momentum and energy
\cite{RHIC_1,RHIC_2,RHIC_3,RHIC_4,FlowSPS,RachidQM2006,PHOflow1,PHOflow2}. 
The collective behavior of
the constituents of created matter is explored by measuring the
particles' azimuthal angular distributions with respect to the
reaction plane. The direction of the reaction plane is defined by the
impact parameter, \bim, of the colliding nuclei and the beam's
direction, Fig.~\ref{fig:VSch}(a). In non-central collisions of
heavy-ions at high energy, the configuration space anisotropy is
converted into momentum space anisotropy, Figs.\ref{fig:VSch}(b)-(c).
The method used to measure the particle flow is based upon the scheme
described by Poskanzer and Voloshin \cite{Posk98} and, considering
that nuclei have some density fluctuations \cite{Alver2010,Han2011},
the strength of the $n^{th}$ flow parameter is given by the $n^{th}$
Fourier coefficient of the particle azimuthal angle distribution
described by the following equation:
\begin{equation}                                                              
\frac{dN}{d(\phi-\psi_{n})}= \frac{1}{2\pi}\big(1+
2\sum_{n=1}^{+\infty} v_{n} cos\ [ n(\phi -\psi_{n})]\big)
\label{eq3}                                                                   
\end{equation} 
here $\phi$ is the azimuthal angle of each hadron. $\psi_{n}$ is the
$n^{th}$ event plane, which varies due to event-by-event
fluctuations~\cite{Alver2010,Han2011}. $\phi$ and $\psi_{n}$ are
determined in the same reference frame (i.e. detector frame).  The
second Fourier expansion (quadrupole component), $v_{2}$, usually is
defined as the strength of the elliptic flow.  \FigureVEngy

The characterization of the collective flow of produced particles by
their azimuthal anisotropy has proven to be one of the more fruitful
probes of the dynamics in \AuAu collisions at RHIC. Elliptic flow,
which is related to the initial spatial shape of the produced matter,
has been of particular interest, as it can provide much information
about the degree of thermalization of the hot, dense
medium. Therefore, it is beneficial to study flow in multiple systems
and compare.  Specifically, \CuCu collisions are interesting as the
copper nucleus is one-third of the size of gold. Exactly how flow
scales between collision systems (e.g. a simple scaling with the
system's size or geometry, the number of valence quarks, or transverse
momentum) is crucial to understand the properties of the produced
matter. The dependence of elliptic flow on the geometry, like \AuAu to
\CuCu of the collision is of particular importance, as flow is thought
to depend heavily on the initial spatial anisotropy. Additionally, any
fluctuations would be expected to have more of an effect in a smaller
system. The unique large pseudorapidity coverage of the PHOBOS
detector ($|\eta| \le$ 5.4) made it ideally suited for probing the
longitudinal structure of the collision, the dynamics of which have
begun to be understood away from mid-rapidity \cite{Hir2006}.

Figure~\ref{fig:VEta} shows the elliptic flow, $v_{2}$, signal as a
function of pseudorapidity, $\eta$, in \AuAu collisions at \snn =
19.6, 62.4, 130 and 200 GeV as well for \CuCu collisions at \sn\ =
22.4, 62.4 and 200 GeV for the 0-40\% most central events. The
resemblance of \AuAu and \CuCu results is striking. The $v_{2}$ in the
smaller system, \CuCu, collisions displays a similar shape in
pseudorapidity and magnitude to that in \AuAu collisions at a given
energy. The strength of the $v_{2}$ signal from \CuCu is surprising in
light of expectations that the smaller size of the system would result
in a much smaller flow signal \cite{Lie2006}.
\FigureVpt

For comparison, Fig.\ref{fig:VEngy} shows the dependence of the
integrated elliptic flow of charged particles, $v_{2}$, on beam energy
at GSI (EOS, FOPI experiments at CERN), AGS (E95 and E877 experiments
at BNL), SPS (CERES and NA49 experiments at CERN), RHIC (PHENIX,
PHOBOS and STAR experiments at BNL), and LHC (ALICE experiment at
CERN)~\cite{FOPI2005,RachidLK2007,Urs2008,Raimand2011_1,Raimand2011_2}.
We observe that at low fixed target energies (\sn $\sim$ 3 GeV),
particle production is enhanced in the direction orthogonal to the
reaction plane, and $v_{2}$ is negative. This is due to the effect
that the spectator parts of the nuclei block the matter in the
direction of the reaction plane. At higher center-of-mass energies,
these spectator components move away sufficiently quickly, and
therefore particle production is enhanced in the reaction plane,
leading $v_{2}$ to be significantly different from zero.  This
phenomenon is expected in hydrodynamic scenarios in which the large
pressure gradients within the reaction plane drive a stronger
expansion. However, the important observation is that, up to the
highest center of mass energies at RHIC, the observed asymmetry
$v_{2}$ continues to grow. The ALICE measurement at 2.76 TeV shows
that the integrated elliptic flow increases by about 30\% compared to
flow measured at the highest RHIC energy of 200 GeV. This result
indicates that the hot and dense matter created in these collisions at
the LHC energy (2.76 TeV) still behaves similar to matter created at
RHIC ~\cite{Raimand2011_1,Raimand2011_2}.  \FigureVnqStar

Figure~\ref{fig:Vpt} presents the dependence of $v_{2}$ on the
transverse momentum (${\rm p_{_{T}}}$) of charged hadrons in \AuAu and
\CuCu collisions at \sn\ = 62.4 and 200 GeV for centrality bins 0-20\% in panels (a)-(b)
and 20-40\% in panels (c)-(d). For both systems, the dependence of $v_{2}$ on \pT is similar to that for the two measured centrality classes, \AuAu and
\CuCu.  For a given system, the $v_2$ shows an increasing trend as a
function of \pT, and there is a significant centrality dependence
where the $v_2$ increases when moving from the most central collisions
(0-20\%) to mid-central ones (20-40\%). This is consistent with what
one would expect from the initial geometry of collisions because the
most central collisions are more circular in shape, therefore a
smaller $v_2$ is expected.

Figure~\ref{fig:VnqStar}(a)-(c) show the anisotropic flow distributions,
$v_{2}$, of identified hadrons (${ \phi}$, ${ \Xi}$, ${
  \Omega}$, ${ \Lambda}$, ${ K^{0}_{S}}$, ${ \pi}$, ${
  K}$ and ${p}$) as a function of the transverse momentum,
\pT. The results were measured for three collision centralities,
0-80\%, \mbox{0-30\%} and 30-80\% in \AuAu collisions at \snn = 200
GeV \cite{Nasim2013}.  In the lower \pT region, \pT\ $\le$ 2 GeV/c,
the value of $v_{2}$ is inversely related to the mass of the hadron,
that is, characteristic of hydrodynamic collective motion in
operation, Fig.~\ref{fig:VnqStar}(a)-(c). At the intermediate
\pT\ region, the dependence is different. Instead of a mass
dependence, there seems to be a hadron type dependence.  We observed a
splitting between baryon and meson $v_{2}$ at intermediate \pT\ for
centrality 0-30\%. However, for 30-80\% centrality, no such distinct
grouping was evident among the baryons and among the mesons.
  
An interesting result is that the multi-strange hadrons exhibit a
 smaller $v_{2}$ than other identified hadrons, i.e. the $\phi$ meson,
 whose mass is close to that of $p$ and ${\Lambda}$, shows \mbox{$v_{2}$
 (${\phi}$) $<$ $v_{2} ({p})$}. It is known that the $\phi$ meson
 does not participate as strongly as others do in hadronic
 interactions, nor can $\phi$ mesons be formed via the
 coalescence-like $K^{+} + K^{-}$ process in high energy collisions
 \cite{StarKprocess}. Hence, the strong $v_{2}({\phi})$ we
 recorded must have developed before hadronization. This result offers
 evidence for partonic collectivity \cite{RHICStar}, which could also
 explain the reduce $v_{2}({\phi})$ in peripheral, as seen in
 Fig.~\ref{fig:VnqStar}(c).  
\FigureVnqPhenix

To include the effect of the collective motion, usually the transverse
momentum of the particle, \pT, is transferred to the transverse
kinetic energy ${\rm KE_{T}}$ $\equiv$ ${\rm m_{T}- m_{0}}$ $=$ ${\rm
  \sqrt{{\it p}_{_{T}}^{2}+m^{2}_{0}}-m_{0}}$, where ${\rm m_{0}}$ is
the particle's mass \cite{Nasim2013}. The measured \vell was scaled by
the number of valence quarks (NVQ) in a given hadron. For mesons and
baryons, respectively, they are \nvq = 2 and 3. The \KET\ also was
scaled with the same ${\rm n_{vq}}$. The results are shown in
Fig.~\ref{fig:VnqStar}(d)-(f). The STAR collaboration observed for the
first time the NVQ scaling in \AuAu at \snn = 200 GeV for 0-80\%
collision centrality, it was considered as a good signature of
partonic collectivity \cite{STAR2005,STAR2008}.  It is interesting to
investigate the NVQ scaling for different centralities that could help
us to understand partonic collectivity for different sized
systems. \mbox{Figure~\ref{fig:VnqStar}(d)-(f)} shows $v_2$ scaled by
\nvq as a function of \KET/\nvq in Au + Au collisions at
\snn~=~200~GeV for the same collision centralities discussed
above. The results in Fig.~\ref{fig:VnqStar}(e) show that scaling
holds for all identified strange hadrons for \mbox{0-30\%}
centrality. This indicates that the major part of flow could be
developed at the partonic phase for \mbox{0-30\%} centrality. On the
other hand, for 30-80\% centrality shown in the
Fig.~\ref{fig:VnqStar}(f), the ${\rm \phi}$-meson deviates by about
10\% from the fit line for the range \mbox{\KET/\nvq $>$ 0.6
  GeV/c}. This could be interpreted as due to the small contribution
from the partonic phase to collectivity.  It should be noted that
current LHC data, \PbPb collisions at \snn = 2.76 TeV, exhibit
deviations from the observed number of valence quark scaling at the
level of about $\pm$ 20\% for \pT $>$ 3 GeV/c \cite{LHVv2nq}.
\FigureVnIP

Measuring the event's anisotropy components, $v_n$, is a powerful
probe for investigating the characteristics of the created medium at
RHIC.  The results of the anisotropy of these higher-order events
constrain the initial geometrical eccentricity and viscosity that are
used in hydrodynamical models. The $v_n$ of inclusive charged hadrons,
where $n =$ 2, 3, 4 and 5 are measured by RHIC experiments as a
function of \pT\ in various centralities for \AuAu collisions at \snn
= 200 GeV \cite{ppg132}, are shown in \mbox{Figs.~\ref{fig:VnqPhenix}
  and \ref{fig:VnIP}}. All $v_n$ exhibit an increasing trend as a
function of \pT\ and $v_2$ is the dominant component. The third
harmonic, or $v_3$, for which the main source is believed to be from
fluctuation~\cite{Alver2010}, shows an interesting behavior. Unlike
$v_2$, the $v_3$ as a function of \pT\ is somewhat independent from
collision centrality, which means the source of $v_3$ probably is not
strongly influenced by the initial collision geometry.  Also
measurements of event-by-event $v_3$ provide additional constraints on
theoretical models.

A good candidate for providing initial conditions for systematic flow
studies is the Impact Parameter dependent Glasma (IP-Glasma) model
described in detail in
Refs. \cite{Char2013,Schen2012A,Schen2012B,Schen2014} and shown in
Fig.\ref{fig:VnIP} as curves. It combines the IP-Sat (impact parameter
saturation) model of high energy nucleon (and nuclear) wave functions
with the classical SU(3) Yang-Mills dynamics of the Glasma fields
produced in a heavy-ion collision.  To describe the flow measurements,
the IP-Glasma model of the classical early time evolution of
boost-invariant configurations of gluon fields was coupled to a
viscous hydrodynamic model to describe the system's evolution
\cite{Char2013}. Based on this comparison presented in
Fig.~\ref{fig:VnIP}, the data are well described by the IP-Glasma
model given the systematic uncertainties in both the experimental and
theoretical calculations \cite{Char2013}. The fact that the IP-Glasma
model coupled with a viscous hydrodynamic model, which assumes the
formation of a QGP, reproduces the experimental data implies that the
fluctuations in the initial state are important and the created medium
behaves as a nearly perfect liquid of nuclear matter because it has an
extraordinarily low ratio of shear viscosity to entropy density,
$\eta/s$ $\approx$ 0.12, see Fig.\ref{fig:VnIP}. It even appears to be
the most perfect fluid observed in nature so far, having a specific
viscosity (viscosity to entropy density ratio) at least an order of
magnitude smaller than that of any previously observed liquid.

\section{Experimental evidence of created dense matter \label{sec:dense}}
\subsection{High-\pT\ hadron suppression: jet-quenching}
 At high momentum, the perfect fluid collectivity is expected to
 breakdown, and in these models viscous corrections become large
 \cite{STARF2005,Gyul2015}.  In Figs.~\ref{fig:VnqStar},
 \ref{fig:VnqPhenix} and \ref {fig:VnIP}, instead of continuing to
 rise with \pT, the elliptic asymmetry stops growing at \pT\ $>$ 2
 GeV/c, and the difference between baryon versus meson, $v_2$ became
 distinct as in Fig.~\ref{fig:VnqStar}(b). In a coalescence picture,
 the quarks contributing to baryons are at lower momentum and so more
 strongly coupled compared to these for meson~\cite{Mol2003}.
 Furthermore, the PHENIX collaboration reported an important result at
 RHIC is that the non-equilibrium power-law high-\pT\ particle
 distributions persist, but are strongly quenched
 \cite{ppg014,ppg024}. Figure~\ref{fig:Spectra} compares the $\pi^0$
 invariant differential cross sections obtained from minimum-bias
 \pp\ collisions at 200 GeV to \AuAu collisions at the same energy and
 for two collision centralities: (a) 0-10\%, and, (b) 80-92\%
 \cite{ppg014,ppg024}. For comparison, the \pp\ data are scaled by
 \Ncoll, corresponding to the centrality in \AuAu collisions. The
 scaled \pp\ data are well parameterized by a power-law form $A$(1
 \pT\ /$p_0$)$^{-n}$, with parameters of $A =$ 393 mb GeV$^{-1}$
 $c^3$, $p_{0}$ = 1.2112 GeV/c, and $n =$ 9.97 as in
 \cite{ppg014,ppg024}. We also observe that the scaled \pp\ data are
 well reproduced by the NLO pQCD calculations
 \cite{ppg014,ppg024}. Figure~\ref{fig:Spectra}(a) shows clearly that
 the $\pi^0$ invariant yield from central (0-10\%) \AuAu collisions as
 a function of \pT\ are suppressed compared to the $\pi^0$ invariant
 yield from \pp\ collisions at the same \pT.  This is clear
 experimental evidence of the quenching phenomena of $\pi^0$ in
 central \AuAu collisions at maximum RHIC energy, \snn\ = 200 GeV. In
 contrast, for the \AuAu peripheral collisions (80-92\%) shown in
 Fig.~\ref{fig:Spectra}(b), the production rate of $\pi^0$ particles
 agrees well with $\pi^0$ particles production in \pp\ collisions and
 in the pQCD theoretical predictions.  \FigureSpectra
  
We quantify the medium effects on high \pT\ production in
nucleus-nucleus collisions, \AAA, with the nuclear modification factor
which is defined as follows:
$${\rm R_{AA} (p_{T})= \frac{ 1 }{\langle N_{coll}\rangle} \times \frac{yield\ per\ A+A\ collision}{
    yield\ per\ {\it p+p}\ collision}}$$
$${\rm = \frac{ 1 }{\langle N_{coll}\rangle} \times
  \frac{d^{2}N^{^{A+A}}/d\eta dp_{T}}{d^{2}N^{^{\it p+p}}/d\eta dp_{T}}}$$

This factor reflects the deviation of measured distributions of
nucleus-nucleus, \AAA, transverse momentum, at given impact parameter
\bim, from measured distributions of an incoherent superposition of
nucleon-nucleon (\pp) transverse momentum (${\rm R_{AA}}$ = 1). This
normalization often is known as ``binary collisions scaling". In the
absence of any modifications due to the `embedding' of elementary
collisions in a nuclear collision, we expect ${\rm R_{AA}}$ = 1 at
high-\pT. At low \pT, where particle production follows a scaling with
the number of participants, the above definition of ${\rm R_{AA}}$
leads to ${\rm R_{AA}}$ $<$ 1 for \pT\ $<$ 2 GeV/c.  \FigureRaaSPSLHC
\FigureRaaRHIC

At maximum SPS energy (\PbPb \snn = 17.3 GeV), the WA98 data for
\RAA\ presents an enhancement of moderately high-\pT\ tails in central
\PbPb collisions \cite{david2004}. This enhancement was anticipated as
a result of the Cronin effect \cite{Cronin75}.  Another experimental
observation in heavy-ion collisions (from AGS to RHIC energies) is
that hadron production at mid-rapidity (${\rm |y| <}$ 1.5) rises
logarithmically with increasing collision energy.  At RHIC, the
central zone is approximately net-baryon-free
\cite{BrahmsBaryon}. Particle production is large and dominated by
pair production, and the energy density seems to exceed significantly
that required for QGP formation \cite{RHIC_1,RHIC_2,RHIC_3,RHIC_4}.

The evolution of the nuclear modification factor with center-of-mass
energy, from the SPS~\cite{david2004} to RHIC
\cite{RachidHDR,CompilationRaa_1,CompilationRaa_2,CompilationRaa_3,CompilationRaa_4,RN2011,NN2012}
and then to the LHC~\cite{CMS}, is presented in
Fig.~\ref{fig:RaaSPSLHC}. We observe that in the presented \pT region,
charged hadron production at the LHC is found to be about 50\% more
suppressed than at RHIC, and has a similar suppression value as for
neutral pions ($\pi^{0}$) measured by PHENIX. These measurements of
\RAA at RHIC and LHC is consistent with a large energy loss in the
medium causing it to become opaque to the propagation of high momentum
partons. In contrast, \RAA for $\pi^{0}$ at SPS energy, as shown
Fig.~\ref{fig:RaaSPSLHC}, exhibit enhancement similar to the \dAu at
RHIC.

RHIC experiments, as shown Fig.~\ref{fig:RaaSPSLHC}, revealed the
suppression of the high-\pT hadron spectra at the mid-rapidity region
in central \AuAu collisions compared to the scaled momentum spectra
from \pp\ collisions at the same energy, \mbox{\snn = 200 GeV.}  This
effect, proposed by Bjorken, Gyulassy, and others
\cite{BjorkenGyulassy_1,BjorkenGyulassy_2} rests on the expectation of
a large energy loss of high momentum partons scattered in the initial
stages of collisions in a medium with a high density of free color
charges. According to the QCD theory, colored objects will lose energy
via bremsstrahlung radiation of gluons \cite{Gaard}.  Such a mechanism
strongly would degrade the energy of leading partons, as reflected in
the reduced transverse momentum of leading particles in the jets
emerging after their fragmentation into hadrons. This energy loss is
expected to depend strongly on the color charge density of the created
system, and the path length traversed by the propagating parton. It
has been recognized that the elastic energy loss is too small to
engender significant attenuation of the jet in the created medium
\cite{Eloss}. The STAR experiment at RHIC established that the
topology of high-\pT\ hadron's emission is consistent with jet
emission, so that jet-suppression is a valid concept and is elaborated
in the next section.

In \dAu collisions at \snn = 200 GeV, see Fig.~\ref{fig:RaaSPSLHC},
the ${\rm R_{dA}}$ for charged hadrons (h$^{+}$$+$h$^{-}$)/2 is
enhanced. This enhancement is an initial state effect \cite{Accardi}
known by Cronin effect which also is seen in \pA collisions. The
Cronin effect is associated with the initial multiple scattering of
high momentum partons.  In contrast, the RHIC results show a factor of
4 to 5 suppression in central \AuAu (at \snn\ = 200 GeV). In the same
context, the ${\rm R_{dAu}}$ of $\pi^{0}$ and $\eta$ in central \dAu
collisions exhibited no suppression of high-\pT, in contrast with the
${\rm R_{AA}}$ of $\pi^{0}$ and $\eta$ in central \AuAu collisions
that is suppressed, as shown in Fig.~\ref{fig:RaaRHIC}(a). At \pT
$\sim$ 4 GeV/c, we find a ratio ${\rm R_{dAu}}$/${\rm R_{AA}}$
$\approx$ 5. Indeed, the ${\rm R_{dA}}$ distribution shows the Cronin
type enhancement observed at lower energies as in \PbPb collisions at
17.3 GeV/c \cite{McGyl2005,david2004} as shown in
Fig.~\ref{fig:RaaSPSLHC}.

Figure~\ref{fig:RaaRHIC}(b) summarizes the ${\rm R_{AA}}$ for direct
photons, ${\rm \pi^{0}}$ and ${\rm \eta}$ in central \AuAu collisions
at \snn = 200 GeV. The ${\rm R_{AA}}$ for direct photons are not
suppressed as they do not interact strongly with the medium. The ${\rm
  R_{AA}}$ for both ${\rm \pi^{0}}$, ${\rm \pi^{\pm}}$ and ${\rm
  \eta}$'s mesons exhibit the same suppression relative to the
point-like scaled \pp data by a factor of $\sim$~5 that appears to be
constant for \pT\ $>$ 4 GeV/c while the ${\rm \eta}$ mass is much
larger than that of ${\rm \pi^{0}}$ (and ${\rm \pi^{\pm}}$). These
observations of high-\pT hadron suppression, combined with the
collectivity results from Fig.~\ref{fig:VnqStar}, suggest that the QCD
medium created at RHIC is both partonic and opaque. Adding to this
discovery of suppression of particles at high transverse momentum, two
very striking results were seen for open heavy flavor from the PHENIX
experiment via the measurement of electrons from the semi-leptonic
decays of hadrons carrying charm or bottom quarks
\cite{CompilationRaa_1,CompilationRaa_2,CompilationRaa_3,CompilationRaa_4}. First,
heavy mesons, despite their large mass, exhibit a suppression at high
transverse momentum compared to that expected from $p+p$ collisions
\cite{CompilationRaa_2,HeavyPHENIX,HeavySTAR}. This suppression is
similar to that of light mesons, which implies a substantial energy
loss of fast heavy quarks while traversing the medium, see
Fig.~\ref{fig:RaaHF}(a). Second, an elliptic flow is observed for
heavy mesons, electrons from semi-leptonic decays of hadrons carrying
charm or bottom quarks, that is comparable to that of light mesons
like pions, Fig.~\ref{fig:RaaHF}(b). This implies that the heavy
quarks in fact are sensitive to the pressure gradients driving
hydrodynamic flow giving new insights into the strongly coupled nature
of the QGP fluid at these temperatures.  \FigureRaaHF
 
These data on heavy mesons ${\rm R_{AA}}$ were described by
theoretical calculations of the parton energy loss in the matter
created in Au+Au collisions as shown in the Fig.~\ref{fig:RaaHF}
\cite{NN2012,Vitev1,Wang2004}. From these theoretical frameworks, we
learned that the gluon density ${\rm dN_{g}}$/dy must be approximately
1000, and the energy density of the matter created in the most central
collisions must be approximately 15 GeV/fm$^{3}$ to account for the
large suppression observed in the data \cite{Vitev2,Vitev3}.

\subsection{Dijet fragment azimuthal correlations: opaque medium}
Hard-scattering processes were established at high-\pT\ in elementary
collisions at high energy through the measurement of jets
\cite{Jet1,Jet2,Jet3}, back-to-back jets (dijets) \cite{Jet4},
high-\pT\ single particles, and back-to-back correlations between
high-\pT\ hadrons \cite{Jet5}. Jets were shown to carry the momentum
of the parent parton \cite{Jet6}. The jet's cross sections and
high-\pT\ single particle spectra are well described over a broad
range of energies in terms of the hadron’s parton distributions, hard
parton scattering treated by pQCD, and subsequent fragmentation of the
parton \cite{Ake83_1,Ake83_2,Ake83_3}. In the absence of the effects
of the nuclear medium, the rate of hard processes should scale
linearly with the number of binary nucleon-nucleon collisions.

Results from RHIC, described above, show a suppression of the single
particle's inclusive spectra of hadrons for \pT\ $>$ 2 GeV/c in
central \AuAu collisions, indicating substantial in-medium
interactions \cite{Adc02002,Adler2002}. The study of high transverse
momentum hadron production in heavy-ion collisions at RHIC provides
the opportunity to probe in more detail the evolution of the matter
produced and experimentally probe the QCD matter in the most dense
stage of the collisions, wherein quark-gluon deconfinement is likely
to occur \cite{Jacobs2005}.  Measurements of two-hadron angular
correlations, as seen in Fig.~\ref{fig:Dijets}, at large transverse
momentum for \pp and \AuAu as well as \dAu collisions at the same
energy, \snn = 200 GeV, provide evidence for the production of jets in
high-energy nucleus-nucleus collisions and allow first measurements,
inaccessible in inclusive spectra, of the fate of back-to-back jets in
the dense medium, thereby serving as a tomography probe of the medium.

The STAR collaboration established the first measurements of
two-particle azimuthal distribution $D (\Delta \phi)$, defined as
\begin{equation}                                                              
D (\Delta \phi) = \frac{1}{N_{trigger}}\frac{1}{\epsilon} \frac{dN}{d(\Delta \phi)}              
\label{eq4}                                                                   
\end{equation} 
at the mid-rapidity region \mbox{$|\eta|$ $<$ 0.7}. $N_{trigger}$
is the number of particles defined in specific \pT\ range, referred to
as trigger particles. $\epsilon$ is the tracking efficiency of the
associated particles \cite{Jetstar,STARdijet_1,STARdijet_2}.

Figure~\ref{fig:Dijets} shows the discovery of mono-jet production
\cite{GyulJet} in central \AuAu collisions at the maximum RHIC
energy. The pedestal-subtracted azimuthal distributions for high-\pT
pairs in \pp and central \dAu collisions show clear di-jet
features. The azimuthal distributions are shown also for central \AuAu
collisions after subtracting the elliptic flow and pedestal
contributions \cite{Jetstar}. The measurements were obtained with a
trigger particle \ptt\ under the condition 4 $<$ \ptt\ $<$ 6 GeV/c,
along with an associated particle with \pta: 1) for 2 $<$ \pta$<$ \ptt
GeV/c as shown in Fig.~\ref{fig:Dijets}(a), and 2) for \mbox{0.15 $<$
  \pta$<$ 4 GeV/c} as shown in Fig.~\ref{fig:Dijets}(b).  The
two-particle correlation is presented as a function of the difference
between the azimuthal angles of the two particles (${\rm \Delta
  \phi}$) produced in the \AuAu, \dAu and \pp\ collisions at the same
energy, \snn~= 200~GeV.  \FigureDijets%

The results illustrated in Fig.~\ref{fig:Dijets}(a) demonstrate that
the trigger-side correlation peak (at $\Delta \phi =$ 0) in central
\AuAu collisions apparently is the same as that measured in \pp\ and
\dAu collisions but the away-side jet correlation (at $\Delta \phi =
\pi $) in \AuAu has vanished. This observation is consistent with a
large energy loss in the medium causing it to become opaque to the
propagation of high momentum partons. However, it is noticeable in
Fig.~\ref{fig:Dijets}(b), which has a wide \pta, 0.15 $<$ \pta $<$ 4
GeV/c, that the jets on the away side have not disappeared, but simply
lost energy so that the away-side correlation peak has become much
wider than that of the \pp\ collisions. These collective measurements,
from \pp, \dAu and \AuAu of two-particle correlations, point to the
creation of a dense medium in central \AuAu collisions at \snn~= 200
GeV.

The main goal was to measure the dijet fragment azimuthal correlations
in \dAu collisions at similar energy as in \AuAu collisions, so as to
check the interpretation of high-\pT suppression in \AuAu at \snn =
200 GeV; quenching was due to initial state saturation (shadowing) of
the gluon distributions and/or to jet quenching at the final
state. One method to asset the QGP final state interactions was by
substituting a deuterium beam for one of the two heavy nuclei of \AuAu
and it has been predicted that such a control test would be essential
to set apart the unknown nuclear gluon shadowing contribution to the
A+A quench pattern \cite{Wang1992}. Furthermore, \dAu was proposed to
test the predictions of the possible initial state Cronin multiple
interactions \cite{Accardi,Wang2000,Vitev2003,Qiu}. In contrast, one
model of the Color Glass Condensate anticipated a substantial
suppression in \dAu collisions at \snn\ = 200 GeV \cite{Khar2003}. The
RHIC results \cite{Shado2003_1,Shado2003_2,Shado2003_3,Shado2003_4}
definitively rule out a large initial shadowing as the cause of the
quenching in \AuAu\ (shown in
Figs.~\ref{fig:RaaSPSLHC}~and~\ref{fig:RaaRHIC}). The return of
back-to-back jet correlation in \dAu to the level observed in \pp is
illustrated in Fig. ~\ref{fig:Dijets}(a). The results turn out
consistent with jet quenching as a final state effect in \AuAu.  These
\dAu results hold the conclusion that the observed jet quenching in
\AuAu is due to the partons' energy loss \cite{Wang2004}. Jet
quenching theoretical analyses support the estimates of energy density
determined from measuring charged particle multiplicity by RHIC
experiments. They estimated large energy losses for jets propagating
through the medium, and enhance the case for multiple strong
interactions of the QGP constituents of the medium created at RHIC
\cite{Gyulassy2004}.
\section{Conclusions}                                                                  
Quantum chromodynamics, QCD, is the base of the Standard Model,
providing a fundamental description of hadron physics in terms of
quark and gluon degrees of freedom. The theory has been checked
broadly, particularly in inclusive and exclusive processes including
collisions at large momentum transfer where factorization theorems and
the smallness of the QCD effective coupling concede perturbative
predictions \cite{Brodsky}. QCD is a very rich and complex theory,
prominent to many new physical phenomena. Ultrarelativistic heavy-ion
collisions offer the unique ability to investigate hot and dense QCD
matter under laboratory conditions.  However, due to the fundamental
confining properties of the physical QCD vacuum, the deconfined quanta
(i.e. the quarks and gluons) are not directly observable. What is
observable are hadronic and leptonic residues of this transient
deconfined state.  Collisions of protons or heavy-ions, which are
accelerated to nearly the speed of light, are an excellent system to
study these interactions in more detail. In contrast to proton-proton
collisions, where only a small number of particles are created, the
particle production in heavy-ion collisions is so enormous that a QCD
medium with collective behavior and unique properties is produced. It
consists of quarks and gluons and is, therefore, called the quark
gluon plasma, QGP. Relativistic heavy-ion collisions are carried out,
for instance, at the LHC at CERN and RHIC at BNL, where experimental
evidence was found that in those collisions a new state of nuclear
matter indeed is produced.

In this review article, we have focused on measurements as a function
of system sizes, collision centrality and energy carried out in RHIC
experiments. We have presented experimental measurements on the
production of bulk particles, the initial conditions of the
collisions, the particle ratios and the chemical freeze-out conditions
for different particle species, as well as the flow anisotropy of
particles, high-p$_{t}$ hadron suppression, dijet fragment azimuthal
correlations, and heavy flavors probes. We compared these experimental
measurements to theoretical model calculations. The measurements
suggest that RHIC discovered a new state of matter in central \AuAu
collisions at \snn = 200 GeV. This new form of matter is hot, dense,
and strongly interacting, and is consistent with the hot, dense state
as predicted from QCD. Precision measurements already are providing
detailed information on its initial conditions and transport
properties. Whilst this material, and its discovery, is far from being
fully understood, the focus of the RHIC experiments, for instance, is
on a detailed exploration of the properties of this new state of
nuclear matter, QGP, by upgrading the detectors and beam luminosities.

In 2010, the LHC entered the field of relativistic heavy-ion physics
with a very impressive performance. With a wealth of results within a
few months after the first collisions, the LHC heavy-ion program has
benefited from many years of preparation, a very strong and
complementary set of detectors, and a decade of experience and
progress made at RHIC. The number of new observations into the
properties of QCD matter under extreme conditions continues to
rise. The conclusion established at RHIC, that a hot, dense medium
that flows with a viscosity-to-entropy density ratio close to the
conjectured lower bound and quenches the energy of hard probes was
confirmed at the LHC.  However, one cannot avoid pointing out that
recent experimental results from LHC and RHIC provide new insight into
the role of initial and final state effects both in \pPb\ and \dAu\
collisions. The latter have proven to be interesting and more
surprising than original anticipated; and could conceivably shed new
light in our understanding of collective behavior in heavy-ion
physics. Therefore, the experiments at RHIC and the LHC are probing
complementary kinematic regions and main goal of the present project
is to attain, at both colliders, an understanding of the properties of
the matter created, QGP, that is believed to be the primordial matter
present in the Early Universe

\section*{Acknowledgements}
The author thanks F.~Karsch, B.~Schenke, P.~Stankus,
C.~Beck, D.~Morrison, M. McCumber and L. McLerran for their stimulating discussions. P.~Tribedy,
R.~Venugopalan and A. Andronic are warmly thanked for useful
suggestions and for providing model calculations plotted on
Figs.~\ref{fig:MultMid}~and~\ref{fig:Ratiotherm}. C.~Roy and Institut
Pluridisciplinaire Hubert Curien (IPHC) as well as A.~Nourreddine and
Universit\'e de Strasbourg are fully acknowledge for their supports
for HDR ("Habilitation \`a Diriger des Recherches") where parts of the
present review paper were presented in the HDR written document
available at:
\\ {\small\url{https://tel.archives-ouvertes.fr/tel-00925262}}.\\ The
author's research was supported by US Department of Energy,
DE-AC02-98CH10886.
\bibliographystyle{epj}
\bibliography{NOUICER_Rachid_EPJP.bbl} 

\begin{thebibliography}{256}
\bibitem{Shuryak801}E.~Shuryak, Phys. Rep. \textbf{61}, 71 (1980).
\bibitem{Shuryak802}L.~McLerran, Rev. Mod. Phys. \textbf{58}, 1021 (1986).
\bibitem{EarlyUniverse}M. Hindmarsh and O. Philipsen Phys. Rev. D \textbf{71}, 08730 (2005).
\bibitem{phase}{\url{http://www.jicfus.jp/en/promotion/pr/mj/guido-cossu/}}
\bibitem{Sza2010}S.~Borsanyi, G. Endrodi, Z. Fodor, A. Jakovac, S, D. Katz, S. Krieg, C. Ratti, K. K. Szabo, 
Journal of High Energy Physics, \textbf{11}, 077 (2010).
\bibitem{Sza2014}S. Borsanyi, Z. Fodor, C. Hoelbling, S. D. Katz, S. Krieg, K. K. Szabo, Phys. Lett. B \textbf{370}, 99 (2014).
\bibitem{HotQCD}A.~Bazavov  A. Bazavov, T. Bhattacharya, C. DeTar, H. Ding, S. Gottlieb, R. Gupta, P. Hegde, U.M. Heller, F. Karsch, E. Laermann {\it et al.},   Phys. Rev. D \textbf{90}, 094503 (2014). 
\bibitem{Greiner1975} H. G. Baumgardt, J. U. Schott, Y. Sakamoto, E. Schopper, H. St\"{o}cker, J. Hofmann, W. Scheid and W. Greiner, Z. Phys. A \textbf{273}, 359 (1975).
\bibitem{Bjorken}J.D. Bjorken, Phys. Rev. D \textbf{27}, 140 (1983). 
\bibitem{RHIC_1}I. Arsene \textit{et al.} (PHENIX Collaboration), Nucl. Phys. A \textbf{757}, 1 (2005).
\bibitem{RHIC_2}B. B. Back \textit{et al.} (PHOBOS Collaboration), Nucl. Phys. A \textbf{757},  28 (2005).
\bibitem{RHIC_3}
J. Adams \textit{et al.} (BRAHMS Collaboration), Nucl. Phys. A \textbf{757},  102 (2005).
\bibitem{RHIC_4}
K. Adcox \textit{et al.} (STAR Collaboration), Nucl. Phys. A \textbf{757},  184 (2005).
\bibitem{AuAuDesign}M. Harrison, T. Ludlam, S. Ozaki, Nuclear Instruments and Methods in Physics Research A \textbf{499}, 235 (2003).
\bibitem{RHICBeam}{\url{http://www.agsrhichome.bnl.gov/RHIC/Runs/}}
\bibitem{ppDesign} I. Alekseev, C. Allgower, M. Bai, Y. Batygin, L. Bozano, K. Brown, G. Bunce, P. Cameron,
E. Courant, S. Erin \textit{et al.}, Series: C-A/AP - Report Number: C-A/AP/455; BNL-97226-2006-IR.
\bibitem{McGyl2005}M. Gyulassy and L. McLerran, Nucl. Phys. A \textbf{750}, 30 (2005).
\bibitem{LHCATLAS2014}G. Aad  \textit{et al.} (ATLAS Collaboration),  Phys. Rev. Lett. \textbf{114}, 072302 (2015).
\bibitem{LHCALICE2014} K. Aamodt \textit{et al.} (ALICE Collaboration), Phys. Lett. B \textbf{734}, 31 (2014).
\bibitem{Kh2010}V. Khachatryan \textit{et al.} (CMS Collaboration), Journal of High Energy Physics, \textbf{09}, 091 (2010).
\bibitem{Aad2013}G. Aad   \textit{et al.} (ATLAS Collaboration), Phys. Rev. Lett. , \textbf{110}, 182302, (2013).
\bibitem{Mil2014_1}B. Abelev \textit{et al.} (ALICE Collaboration), Phys. Lett. B \textbf{726}, 164 (2013).
\bibitem{Mil2014_2}L. Milano (ALICE Collaboration), J. Phys. Conf. Ser. \textbf{509}, 012105 (2014).
\bibitem{Adare2007} A. Adare \textit{et al.}  (PHENIX Collaboration), Phys. Rev. Lett. \textbf{98} 232301 (2007).
\bibitem{Regeneration_1}X. Zhao and R. Rapp, Phys. Lett. B \textbf{664},
  253 (2008).
\bibitem{Regeneration_2} Y. Liu, Q. Zhen, N. Xu and P. Zhuang, Phys. Lett. B  \textbf{678}, 72 (2009).
\bibitem{Vogt2005}R. Vogt, Phys. Rev. C \textbf{71}, 054902 (2005). 
\bibitem{Khar1997}D. Kharzeev, C. Lourenco, M. Nardi, and H. Satz, Z. Phys. C \textbf{74}, 307 (1997).
\bibitem{Eskola2009}K. J. Eskola, H. Paukkunen, and C. A. Salgado, JHEP \textbf{04}, 065 (2009).
\bibitem{Nagle2011}J. L. Nagle, A. D. Frawley, L. A. Linden Levy, and M. G. Wysocki, Phys. Rev. C \textbf{84}, 044911 (2011).
\bibitem{Rachid2003}R. Nouicer  \textit{et al.} (PHOBOS Collaboration), Eur. Phys. J. C \textbf{33}, S606 (2004).
\bibitem{Ahmad2013}S. Ahmad, A. Ahmad, A. Chandra, M. Zafar and M. Irfan, 
Advances in High Energy Physics, \textbf{2013}, 836071 (2013).
\bibitem{RachidMoriond2002}R. Nouicer {\it et al.} (PHOBOS Collaboration), published in 
2002 QCD and Hadronic Interactions, edited by Tran Thanh Van (The Gioi Publishers,  Hanoi, 2002), pp. 381.
\bibitem{Glauber1970}R. J. Glauber, G. Matthiae, Nucl. Phys. B \textbf{21}, 135 (1970).
\bibitem{Shukla2003}P. Shukla, Phys. Rev. C \textbf{67}, 054607 (2003).
\bibitem{Mich2007}M. L. Miller, K. Reygers, S. J. Sanders, P. Steinberg, Annu. Rev. Nucl. Part. Sci. \textbf{57}, 205 (2007).
\bibitem{Shou2015}Q. Y. Shou, Y. G. Ma, P. Sorensen, A. H. Tang, F. Videbæk and H. Wang,  Physics Letters B \textbf{749}, 215 (2015).  
\bibitem{RachidHDR}R. Nouicer, ``Habilitation \`a Diriger des
  Recherches'', University of Strasbourg, 2013 HDR/N$^{o}$~293\\
{\small\url{https://tel.archives-ouvertes.fr/tel-00925262}}
\bibitem{AuAufrag}B. B. Back {\it et al.} (PHOBOS Collaboration), Phys. Rev. Lett. \textbf{91}, 502303 (2003).
\bibitem{Panic2006}R. Nouicer {\it et al.} (PHOBOS Collaboration), AIP Conf. Proc. \textbf{842}, 86 (2006). 
\bibitem{CuCu2008}B. Alver \textit{et al.} (PHOBOS Collaboration), Phys. Rev. Lett. \textbf{102}, 142301 (2009). 
\bibitem{SPSdNdEta1}S. V. Afanasiev \textit{et al.} (NA49 Collaboration), Phys. Rev. C \textbf{66}, 054902 (2002).
\bibitem{SPSdNdEta2}T. Anticic \textit{et al.} (NA49 Collaboration), Phys. Rev. C \textbf{69}, 024902 (2004).
\bibitem{ALICEPb276}K. Aamodt \textit{et al.} (ALICE Collaboration), Phys. Rev. Lett. \textbf{105}, 252301 (2010).  
\bibitem{IPCGC2012}P. Tribedy, R. Venugopalan, Phys. Lett. B \textbf{710}, 125 (2012).
\bibitem{Dima2001}D. Kharzeev, M. Nardi, Phys. Lett. B \textbf{507}, 121 (2001).
\bibitem{Kov2015}A. Kovner and M. Lublinsky, Phys. Rev. D \textbf{92}, 034016 (2015).
\bibitem{RachidQM2004}R. Nouicer {\it et al.} (PHOBOS Collaboration),  J. Phys. G \textbf{30}, S113 (2004).
\bibitem{dAuPRL}B.B. Back {\it et al.} (PHOBOS Collaboration), Phys. Rev. Lett. \textbf{93}, 082301 (2004).
\bibitem{dAuPRC}B.B. Back {\it et al.} (PHOBOS Collaboration), Phys. Rev. C \textbf{72}, 031901(R) (2005). 
\bibitem{PHOBIG}B.B. Back {\it et al.} (PHOBOS Collaboration), Phys. Rev. C \textbf{83}, 024913 (2011). 
\bibitem{RachidNIM}R. Nouicer {\it et al.} (PHOBOS Collaboration), Nucl. Instrum. Meth. in Physics Research A \textbf{461}, 143 (2001). 
\bibitem{PHONIM} B.B. Back {\it et al.} (PHOBOS Collaboration), Nucl. Instrum. Meth. in Physics Research A \textbf{499}, 603 (2003). 
\bibitem{Thomas}T.S. Ullrich, Eur. Phys. J. A \textbf{19}, s01 (2004). 
\bibitem{Hagedorn1965}R. Hagedorm, Supl. A. Nuvo Cimento Vol III \textbf{No.2}, 150 (1965).
\bibitem{BRAHMSRatio}I. G. Bearden  \textit{et al.} (BRAHMS Collaboration), Phys. Rev. Lett. \textbf{90}, 102301 (2003). 

\bibitem{Anton2013}A. Andronic \textit{et al.}, Nucl. Phys. A \textbf{904-905}, 535c (2013). 
\bibitem{BRAHMSStopping}I. G. Bearden \textit{et al.} (BRAHMS Collaboration), Phys. Rev. Lett.  \textbf{93}, 102301  (2004). 
\bibitem{StarPRC2011}M.M. Aggarwal  \textit{et al.} (STAR Collaboration), Phys. Rev. C \textbf{83}, 034910  (2011). 
\bibitem{Stock2006}R. Stock, Proceedings of Science (PoS CPOD2006): \textbf{040} (2006).
\bibitem{Munz2004}P. Braun-Munzinger, K. Redlich, and J. Stachel, in Quark-Gluon Plasma 3, eds. R. C. Hwa, X. N. Wang (World Scientific, Singapore, 2004), p. 491-599.
\bibitem{Klim2010}M. Kliemant, R. Sahoo, T. Schuster, and R. Stock,  
The Physics of the Quark-Gluon Plasma, Lecture Notes in Physics Volume \textbf{785}, 23 (2010). 
\bibitem{Beca2002}F. Becattini, Nucl. Phys. A \textbf{702}, 336 (2002).

\bibitem{Kars200}F. Karsch, E. Laermann, and A. Peikert, Phys. Lett. B \textbf{478}, 447 (2000). 
\bibitem{Ulri2006} U. Heinz, and G. Kestin, Proceedings of Science (PoS CPOD2006) : \textbf{038} 2006.
\bibitem{Stok79} H. St\"{o}cker, J.A. Maruhn, and W. Greiner, Phys. Rev. Lett. \textbf{44}, 725 (1980).
\bibitem{Stok82} H. St\"{o}cker, LP Csernai, G. Graebner, G. Buchwald, H. Kruse, RY Cusson, 
J. A. Maruhn,  and W. Greiner, Phys. Rev. C \textbf{25}, 1873 (1982).
\bibitem{Stok86}H.St\"{o}cker, and W.Greiner. {\it et al.}, Phys. Rep. \textbf{137}, 277 (1986). 
\bibitem{Reis77}W. Reisdorf and H. G. Ritter, Annu. Rev. Nucl. Part. Sci. \textbf{47}, 663 (1997).
\bibitem{FlowSPS}C. Alt \textit{et al.} (NA49 Collaboration), Phys. Rev. C \textbf{68}, 034903 (2003).
\bibitem{RachidQM2006}R. Nouicer {\it et al.} (PHOBOS Collaboration), J. Phys. G \textbf{34}, S887 (2007).
\bibitem{PHOflow1}B. Alver {\it et al.} (PHOBOS Collaboration), Phys. Rev. Lett. \textbf{94}, 122303 (2005).
\bibitem{PHOflow2}B. Alver {\it et al.} (PHOBOS Collaboration),  Phys. Rev. Lett. \textbf{98}, 242302 (2007). 
\bibitem{Posk98}A. M. Poskanzer, and S. A. Voloshin, Phys. Rev. C \textbf{58}, 1671 (1998).
\bibitem{Alver2010}B. Alver, and G Roland, Phys. Rev. C \textbf{81}, 054905 (2010). 
\bibitem{Han2011}L .X. Han, G. L. Ma, Y. G. Ma, X. Z. Cai, J. H. Chen, and S. Zhang, Phys. Rev. C \textbf{84},  064907 (2011).
\bibitem{Hir2006}T. Hirano, U. Heinz, D. Kharzeev, R. Lacey, and Y. Nara, Phys. Lett. B \textbf{636}, 299 (2006). 
\bibitem{Lie2006}Lie-Wen Chen, and Che Ming Ko, Phys. Lett. B \textbf{634}, 205 (2006).
\bibitem{FOPI2005}  A. Andronic \textit{et al.}  (FOPI Collaboration), Phys. Lett. B  \textbf{612}, 173 (2005).
\bibitem{RachidLK2007}R. Nouicer  \textit{et al.} (PHOBOS Collaboration), Word Scientific, Proceeding of the 22$^{\rm nd}$ Lake Louise Winter Institute, Fundamental Interactions, 373 (2007).
\bibitem{Urs2008}N. Borghini and U. A. Wiedemann, J. Phys. G \textbf{35}, 023001 (2008).
\bibitem{Raimand2011_1}K. Aamodt \textit{et al.} (ALICE Collaboration), Phys. Rev. Lett. \textbf{105} 252302 (2010).
 \bibitem{Raimand2011_2}R. Snellings \textit{et al.} (ALICE Collaboration), J. Phys. G: Nucl. Part. Phys. \textbf{38}, 124013 (2011). 
\bibitem{Nasim2013}Md. Nasim {\it et al.} (STAR Collaboration), Nucl. Phys. A \textbf{904}, 413c (2013).
\bibitem{StarKprocess}J. Adams {\it et al.} Phys. Lett. B \textbf{612}, 181 (2005). 
\bibitem{RHICStar}J. Adams \textit{et al.} Nucl. Phys. A \textbf{757}, 102 (2005).
\bibitem{STAR2005}J. Adams {\it et al.} (STAR Collabortion), Phys. Rev. Lett. \textbf{95}, 122301 (2005).
\bibitem{STAR2008}J. Adams {\it et al.} (STAR Collabortion), Phys. Rev. C \textbf{77}, 054901 (2008). 
\bibitem{LHVv2nq}K. Aamodt \textit{et al.} (ALICE Collaboration),  Journal of High Energy Physics \textbf{06}, 190 (2015).
\bibitem{ppg132}S.S. Adler {\it et al.} (PHENIX Collaboration), Phys. Rev. C. \textbf{89}, 034915 (2014).
\bibitem{Char2013}C. Gale, S. Jeon, B. Schenke, P. Tribedy, and R. Venugopalan, Phys. Rev. Lett. \textbf{110},  012302 (2013).
\bibitem{Schen2012A}B. Schenke, P. Tribedy, and R. Venugopalan, Phys. Rev. Lett. \textbf{108}, 252301 (2012).
\bibitem{Schen2012B}B. Schenke, P. Tribedy, and R. Venugopalan, Phys. Rev. C \textbf{86}, 034908 (2012).
\bibitem{Schen2014}B. Schenke and R. Venugopalan, Phys. Rev. Lett. \textbf{113}, 102301 (2014)
\bibitem{STARF2005}J. Adams et al. (STAR Collaboration), Phys. Rev. C \textbf{72}, 014904 (2005).
\bibitem{Gyul2015}J. Xu, J. Liao and M. Gyulassy, Chin. Phys. Lett. \textbf{32}, 092501 (2015).
\bibitem{Mol2003}D. Molnar, S. A. Voloshin, Phys. Rev. Lett.  \textbf{91},  092301 (2003).
\bibitem{ppg014}S.S. Adler {\it et al.} (PHENIX Collaboration), Phys. Rev. Lett. \textbf{91}, 072301 (2003).
\bibitem{ppg024}S.S. Adler {\it et al.} (PHENIX Collaboration), Phys. Rev. Lett. \textbf{91}, 241803 (2003).
\bibitem{david2004}D. d'Enterria, Phys. Lett. B \textbf{596}, 32 (2004).
\bibitem{Cronin75}
J. W. Cronin, H. J. Frisch, M. J. Shochet, J. P. Boymond, P. A. Pirou\'e, and R. L. Sumner, Phys. Rev. D \textbf{11}, 3105 (1975).
\bibitem{BrahmsBaryon}I.G. Bearden {\it et al.} (BRAHMS Collaboration),  Phys. Rev. Lett. \textbf{93}, 102301 (2004).
\bibitem{CompilationRaa_1}
B.I. Abelev {\it et al.} (STAR Collaboration), Phys. Lett. B \textbf{655}, 104 (2007).
\bibitem{CompilationRaa_2}S.S. Adler  {\it et al.} (PHENIX Collaboration), Phys. Rev. Lett. \textbf{98}, 172302 (2007).           
\bibitem{CompilationRaa_3}S.S. Adler  {\it et al.} (PHENIX Collaboration), Phys. Rev. Lett. \textbf{96}, 202301 (2006).             
\bibitem{CompilationRaa_4}S.S. Adler  {\it et al.} (PHENIX Collaboration), Phys. Rev. Lett. \textbf{91}, 072303 (2003). 
\bibitem{RN2011}R. Nouicer {\it et al.} (PHENIX Collaboration), Nucl. Phys. A {\bf 862}, 62 (2011).  
\bibitem{NN2012}R. Nouicer \textit{et al.} (PHENIX Collaboration),  Journal of Physics : Conference Series \textbf{420},  012021 (2013).
\bibitem{CMS}CMS Collaboration, Eur. Phys. J. C \textbf{72}, 1945 (2012).
\bibitem{BjorkenGyulassy_1}J. D. Bjorken, Phys. Rev. D \textbf{27}, 140 (1983).                    
\bibitem{BjorkenGyulassy_2}M. Gyulassy {\it et al.} Phys. Lett. B \textbf{243}, 432 (1990).   
\bibitem{Gaard} J. J. Gaardhoje {\it et al.} (BRAHMS Collaboration), Nucl. Phys. A \textbf{734}, 13 (2004).
\bibitem{Eloss}M. H. Thoma, and M. Gyulassy \textit{et al.}, Nucl. Phys. B \textbf{351}, 491 (1991).
\bibitem{Accardi} A. Accardi and N. Armesto  \textit{et al.}, Contribution to the CERN Yellow
  report on Hard Probes in Heavy Ion Collisions at the LHC (2002).
\bibitem{HeavyPHENIX}S.S. Alder {\it et al.} (PHENIX Collaboration), Phys. Rev. Lett.  \textbf{96}, 032301 (2006).
\bibitem{HeavySTAR}J. Adams {\it et al.} (STAR Collaboration) , Phys. Rev. Lett. \textbf{94}, 062301 (2005).  
\bibitem{Vitev1}I. Vitev, and Miklos Gyulassy, Nucl. Phys. A \textbf{715}, 779 (2003).
\bibitem{Wang2004}X. N. Wang, Phys. Lett. B \textbf{595}, 165 (2004).
\bibitem{Vitev2}I. Vitev, and M. Gyulassy, Phys. Rev. Lett. \textbf{89}, 252301 (2002).
\bibitem{Vitev3}I. Vitev, J. Phys. G \textbf{30}, S791 (2004).                  
\bibitem{Jet1}M. Banner \textit{et al.} (UA2 Collaboration), Phys. Lett. B \textbf{118}, 203 (1982).
\bibitem{Jet2}G. Arnison \textit{et al.} (UA1 Collaboration),  Phys. Lett. B \textbf{123}, 115 (1983).
\bibitem{Jet3}F. Abe \textit{ et al.} (CDF Collaboration), Phys. Rev. Lett. \textbf{62}, 613 (1989).
\bibitem{Jet4}F. Abe \textit{et al.} (CDF Collaboration), Phys. Rev. D \textbf{41}, 1722 (1990).
\bibitem{Jet5}G. Arnison \textit{et al.} (UA1 Collaboration), Phys. Lett. B \textbf{118}, 173 (1982).
\bibitem{Jet6}F. Abe \textit{et al.} (CDF Collaboration), Phys. Rev. Lett. \textbf{65}, 968 (1990).
\bibitem{Ake83_1}T. Akesson \textit{et al.} (Axial Field Spectrometer Collaboration), Phys. Lett. B \textbf{123}, 133 (1983). 
\bibitem{Ake83_2}J. A. Appel  \textit{et al.} (UA2 Collaboration), Phys. Lett. B \textbf{160}, 349 (1985). 
\bibitem{Ake83_3} G. Arnison  \textit{et al.}  (UA1 Collaboration), Phys. Lett. B \textbf{172}, 461 (1986). 
\bibitem{Adc02002}K. Adcox {\it et al.} (PHENIX Collaboration), Phys. Rev. Lett. \textbf{88}, 022301 (2002).
\bibitem{Adler2002}C. Adler {\it et al.} (STAR Collaboration), Phys. Rev. Lett. \textbf{89}, 202301 (2002).
\bibitem{Jacobs2005}P. Jacobs {\it et al.}  (STAR Collaboration), Prog. Nucl. Phys. \textbf{54}, 443 (2005). 
\bibitem{Jetstar}C. Adler {\it et al.} (STAR Collaboration), Phys. Rev. Lett. \textbf{90}, 082302 (2003). 
\bibitem{STARdijet_1}J. Adams {\it et al.} (STAR Collaboration), Phys. Rev. Lett. \textbf{91}, 072304 (2003).
\bibitem{STARdijet_2}J. Adams {\it et al.} (STAR Collaboration), Phys. Rev. Lett. \textbf{95}, 152301 (2005). 
\bibitem{GyulJet}M. Gyulassy, and M. Pl\"{u}mer, Nucl. Phys. A \textbf{527}, 641 (1991).
\bibitem{Wang1992}X. Wang, and M. Gyulassy,, Phys. Rev. Lett. \textbf{68}, 1480 (1992).
\bibitem{Wang2000}X. N. Wang, Phys. Rev. C \textbf{61},  064910 (2000).
\bibitem{Vitev2003}I. Vitev, Phys. Lett. B \textbf{562}, 36 (2003).
\bibitem{Qiu} J.w. Qiu, and I. Vitev, Phys. Rev. Lett. \textbf{93}, 262301 (2004).
\bibitem{Khar2003}D. Kharzeev, E. Levin, and L. McLerran, Phys. Lett. B \textbf{561}, 93 (2003).
\bibitem{Shado2003_1}B.B. Back {\it et al.} (PHOBOS Collaboration), Phys. Rev. Lett. \textbf{91},  072302 (2003).
\bibitem{Shado2003_2}S.S. Adler {\it et al.} (PHENIX Collaboration), Phys. Rev. Lett. \textbf{91}, 072303 (2003).
\bibitem{Shado2003_3}J. Adams {\it et al.} (STAR Collaboration), Phys. Rev. Lett. \textbf{91}, 072304 (2003).
\bibitem{Shado2003_4}I. Arsene  {\it et al.} (BRAHMS Collaboration), Phys. Rev. Lett. \textbf{91}, 072305 (2003).
\bibitem{Gyulassy2004}M. Gyulassy, I. Vitev, X.N. Wang and B.W. Zhang,
  Quark Gluon Plasma \textbf{3, Eds.} R.C. Hwa and X.-N. Wang, World
  Scientific, Singapore, p. 123. (2004).
\bibitem{Brodsky}S. J. Brodsky, SLAC-PUB-9022.
\end{thebibliography}
\end{document}